\documentclass[aps,prd,twocolumn,amsmath,amsfonts,eqsecnum,nofootinbib,showpacs,showkeys]{revtex4}
\usepackage[pdftex,breaklinks=true,bookmarksnumbered]{hyperref}
\usepackage{graphics}

\newcommand{\vct}[1]{\mathbf{#1}}

\def\nl{\\ & \quad}
\def\nlq{\\ & \quad \qquad}

\def\pa{\partial}

\DeclareMathOperator{\Order}{\mathcal{O}}

\allowdisplaybreaks

\begin{document}

% UFT8 version, for PRD
%\begin{CJK*}{UTF8}{gbsn}

\title{Canonical formulation of gravitating spinning objects at 3.5 post-Newtonian order}

\author{Jan Steinhoff}
\email{jan.steinhoff@uni-jena.de}
\homepage{http://www.tpi.uni-jena.de/gravity/People/steinhoff/}

% UFT8 version, for PRD
%\author{Han Wang (王涵)}
% without CJK, for arXiv
\author{Han Wang (\includegraphics{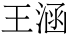})}
\email{han.wang@uni-jena.de}
\homepage{http://www.tpi.uni-jena.de/gravity/People/wang/}

\affiliation{Theoretisch--Physikalisches Institut, Friedrich--Schiller--Universit\"at, Max--Wien--Platz 1, 07743 Jena, Germany, EU}

\date{\today}

\begin{abstract}
The 3.5 post-Newtonian (PN) order is tackled by extending the
canonical formalism of Arnowitt, Deser, and Misner to spinning objects.
This extension is constructed order by order in the PN setting by utilizing
the global Poincar\'e invariance as the important consistency condition.
The formalism is valid to linear order in the single spin variables.
Agreement with a recent action approach is found. A general formula for
the interaction Hamiltonian between matter and transverse-traceless part
of the metric at 3.5PN is derived. The wave equation resulting from this Hamiltonian is
considered in the case of the constructed formalism for spinning objects.
Agreement with the Einstein equations is found in this case.
The energy flux at the spin-orbit level is computed.
\end{abstract}


\maketitle

%\end{CJK*}

\section{Introduction\label{sec:introduction}}
Several laser interferometric gravitational-wave (GW) detectors,
including LIGO, VIRGO, GEO600, and TAMA300 are currently searching for
the GWs emitted by inspiraling compact binaries, which consist of
black holes and/or neutron stars. Because the data analysis method
used by these experiments, namely, the matched filtering technique,
requires the detector's output signal to be compared with a large
amount of theoretical waveforms (templates), the post-Newtonian (PN)
calculation for both of the binary's motion and the gravitational waves
emitted has to be performed. For \emph{nonspinning} binaries, the
PN expansion has been successfully carried out through
3.5PN order. However, since astrophysical observations
suggest that most astrophysical objects carry a certain amount of spin
angular momentum, and compact objects like black holes are usually
rapidly rotating, the effect of spin is too large to be ignored.

Both the dynamics of a \emph{spinning} binary and the GWs emitted by
such a system are very different from those of a \emph{nonspinning}
system. The coupling between the orbital angular momentum ${\bf L}$
and the individual spins ${\bf S}_a$ leads to precession of the
individual spins and the orbital plane, which in turn leads to
additional amplitude modulation of the GWs emitted by the system. The
detailed calculation by Kidder at 1PN order \cite{kidder95} showed
that spin itself can also directly contribute to the gravitational
waveforms and the emission of energy and angular momentum. Including
spin as an intrinsic parameter of the source also increases the
dimension of the parameter space to be used in the data analysis
process, which not only requires more computational resources but
could also affect the accuracy on parameter estimation
\cite{bbw04spin, langhughes06}. Thus, it is desired to carry the
PN approximation for \emph{spinning} binary systems to a
sufficiently high PN order.

The spin effect to the motion and to the gravitational field have been
a long-standing problem in general relativity (GR). Papapetrou and
Corinaldesi in the 1950s \cite{papapetrou51a, papapetrou51b}
calculated the leading order spin effects to the motion of a spinning
test body in a given gravitational field. The leading order spin-orbit
(SO) and spin(1)-spin(2) (S$_1$S$_2$) contributions to the equations of
motion for a system of two spinning black holes were derived by D'Eath
\cite{DEath}, Barker and O'Connell \cite{barker75:so,barker79:ss} in
the 1970s, and later by various authors (for references and reviews,
see, e.g., \cite{threehundred,thornehartle}). Using Thorne's multipole
expansion formalism \cite{thorne80} in terms of symmetric trace-free
(STF) radiative multipoles in the harmonic gauge, the group of Kidder,
Will, and Wiseman derived the leading order SO and S$_1$S$_2$ contribution to
the gravitational radiation flux \cite{kidder92,kidder95} for a
general spinning binary system and the polarized gravitational
waveform emitted by a spinning binary system with quasicircular orbit
\cite{kidder95}. The leading order SO and S$_1$S$_2$
radiation reaction effects to the equations of motion (EOM)
%(which is at 3.5PN order for our formal counting, see below)
%but is of order 4PN for SO and 4.5PN for S$_1$S$_2$ for maximally rotating binary black holes,
were derived by Wang and Will in
the harmonic gauge \cite{dire3,dire4}. A general form of the
SO contributions in arbitrary coordinates was also derived by
Zeng and Will using an energy and angular momentum balance approach
\cite{zengwill07}.

The (conservative) next-to-leading order (NLO) spin effects were only tackled
recently. The first derivation of the NLO SO EOM was
attempted by Tagoshi, Ohashi, and Owen
\cite{tagoshi01:spin}, their result was essentially confirmed
by Faye, Blanchet, and Buonanno in the harmonic gauge
\cite{blanchet:2pnso1}, and later also by Damour, Jaranowski, and Sch{\"a}fer
using a Hamiltonian approach in the Arnowitt, Deser, and Misner
(ADM) gauge \cite{djs08:so2pn} (see also \cite{steinhoff08:spin2}).
The corresponding energy flux (and formulas for a phasing) is given in \cite{blanchet:2pnso2}.
The first attempt to compute the NLO S$_1$S$_2$ contributions to the EOM of
a spinning binary system was made by Porto and Rothstein
\cite{portorafael:2PNspinspin} using an effective field theory
technique, namely, an extension of nonrelativistic general relativity \cite{GR06}
to systems with spin \cite{P06}.
%, but their result was not complete \cite{steinhoff07:spin1}.
The first complete NLO S$_1$S$_2$ Hamiltonian was
presented by Steinhoff, Sch{\"a}fer, and Hergt
\cite{steinhoff07:spin1,steinhoff08:spin2}, and agrees with \cite{PR08a,L08}.

Though the above mentioned results are useful for the creation
of templates, further work needs to be done. In general,
a parametrization of the orbits must be obtained by solving the
EOM. It is common to describe the conservative dynamics
in terms of certain orbital elements, see, e.g., \cite{Wex:1995}.
Spin precession and dissipative effects
can then be described by secular EOM of the orbital elements.
For explicit solutions including spin see \cite{Konigsdorffer:Gopakumar:2005,Tessmer:2009}.
It is also possible to obtain the dissipative orders of these secular EOM
with the help of the conservative parts as well as the energy and angular
momentum flux. In this way secular EOM corresponding to the LO radiation-reaction
EOM mentioned above have already been obtained in \cite{Rieth:Schafer:1997,Gergely:Perjes:Vasuth:1998,Gergely:1999,Gergely:2000}.

In this paper, we extend the canonical formalism of ADM \cite{Arnowitt:1962hi,Regge:Teitelboim:1974,DeWitt:1967} to $n$-body systems
with $n$ spinning objects up to a PN order sufficient for the computation of the SO and
S$_1$S$_2$ contribution to the next-to-next-to-leading order (NNLO)
conservative Hamiltonian $H_{\le 3PN}^{con}$ and the leading order
dissipative Hamiltonian $H_{\le 3.5PN}^{diss}$.
It is important to mention that we count PN orders in a rather formal way;
see Appendix \ref{Scount}.
The canonical framework in the present paper is constructed order by order in
the PN setting by utilizing the global Poincar\'e invariance as the important
consistency condition, similar to \cite{steinhoff08:spin2}.
Further a general formula for the interaction Hamiltonian between matter
and the transverse-traceless part of the metric $h^{\text{TT}}_{ij}$ at 3.5PN is derived.
From this Hamiltonian a wave equation for $h^{\text{TT}}_{ij}$ can be
obtained by canonical methods. For the canonical formalism
presented in this paper, this wave equation agrees with a corresponding
one that can be followed directly from the Einstein equations.
This provides a thorough check of the canonical formalism. Further
the obtained formulas are the basis for applications.
Using the wave equation, we are able to derive, in the radiation zone of a system with
two spinning objects, the leading order SO contribution to
$h^{\text{TT}}_{ij}$ and the energy flux (see also \cite{kidder95,Majar:Vasuth:2008}).

% It is shown that the source of the Hamilton constraint at 3.5PN correctly contains
% information of the projected stress-energy tensor in the ADM gauge, $T_{ij}$, at the 1PN order,
% which enters the source expressions of the wave equation for the transverse-traceless gravitational field $h^{\text{TT}}_{ij}$.

A canonical framework for spinning test-particles valid to any PN order
and linear in the spin of the particle was
given very recently in \cite{Barausse:Racine:Buonanno:2009}. The Hamiltonian
of a spinning test-particle in Kerr spacetime was given explicitly.
This includes parts of the conservative Hamiltonian $H_{\le 3PN}^{con}$
mentioned above (as well as contributions of cubic and higher order in spin;
see also \cite{Hergt:Schafer:2008:2,HS08}).
An action approach to the canonical formulation of self-gravitating
spinning objects valid to all orders linear in the single spin variables
was recently given in \cite{SSepl} and is shown to agree with the present paper up to 3.5PN.
The order by order construction performed in this paper gives
an independent derivation of the results in \cite{SSepl} up to 3.5PN.
Further the method developed in this paper to construct a canonical formalism
might have some advantages over an action approach at higher orders in spin.
Knowledge of the formalism in \cite{steinhoff08:spin2} and its
extension given in the present paper was important information
to succeed with the action approach in \cite{SSepl}.
The consistency checks and applications performed in this paper also
apply to the action approach. Further, the present paper provides more
details than \cite{SSepl}. In particular, the source terms of the
constraints depending on canonical variables are given explicitly.

The paper is organized as follows. In Sec.\ \ref{Sham} it is shown
how the ADM formalism can be extended to spinning objects
order by order in a PN setting. In Sec.\ \ref{Ssource} this
approach is applied to 3.5PN and linear in the spin. A comparison
with the action approach in \cite{SSepl} is given.
Section \ref{SPN} gives the PN expansion of the constraints,
including the matter source terms in canonical variables.
In Sec.\ \ref{Sevo} a general formula for the interaction Hamiltonian
is derived. Further it is shown that the evolution equations
given by this Hamiltonian, specialized to our canonical formalism,
coincide with the ones following from the Einstein equations.
In Sec.\ \ref{Sflux} the SO energy flux is computed.
Section \ref{Sout} gives conclusions and outlook.

Our units are $c=1$ and $G=1$, where $G$ is the Newtonian
gravitational constant. Greek indices will run over $0,1,2,3$, Latin indices
from the middle of the alphabet over
$1,2,3$. Latin indices from the beginning of the alphabet label
the individual objects.
For the signature of spacetime we choose +2. The short-cut notation
$ab$ ($= a^{\mu}b_{\mu} = a_{\mu}b^{\mu}$) for the scalar product of two vectors
$a^{\mu}$ and $b^{\mu}$ will be used. Square brackets denote index antisymmetrization
and round brackets index symmetrization,
i.e., $a^{( \mu} b^{\nu )} = \frac{1}{2} (a^{\mu} b^{\nu} + a^{\nu} b^{\mu})$.
The spatial part of a 4-vector $x$ is $\vct{x}$.
Round brackets around an index denote a local basis, while
round brackets around a number denote the formal order in $c^{-1}$,
as in \cite{steinhoff08:spin2}.

\section{From ADM energy to ADM Hamiltonian\label{Sham}}
In this section we outline how the ADM canonical formalism
\cite{Arnowitt:1962hi,Regge:Teitelboim:1974,DeWitt:1967}
can be extended to self-gravitating spinning objects
order by order in a PN setting. We only consider
a fully reduced canonical framework here where the gauge is fixed
and all constraints are eliminated. Then the ADM energy can be
used as a Hamiltonian, the ADM Hamiltonian, if it is expressed
in terms of variables with standard canonical meaning. The transformation
to such variables can be found from consistency considerations.
At the 3.5PN SO and S$_1$S$_2$ orders the global Poincar\'e algebra
and the constant Euclidean length of the canonical spin uniquely fixes
this transformation to standard canonical variables.
%Further considerations are given in Appendix \ref{poincare}.

The approach outlined here is a natural generalization of the one in \cite{steinhoff08:spin2}.
It was suggested in Appendix B of \cite{steinhoff08:spin2},
by considering the algebra of the gravitational constraints,
that at higher orders spin corrections to the canonical field momentum might be necessary,
and that the gauge structure needs to be extended. Indeed,
the former is an important ingredient of the approach in this
paper [see Eq.\ (\ref{pican})] as well as of the action approach in \cite{SSepl}.
Further, the action approach is based on tetrad gravity, which has
more gauge freedom than metric gravity. In this paper, however,
the gauge is always fixed and the original gauge structure is
less important.

\subsection{Field constraints}
Most important for an explicit calculation of the generators
of the global Poincar\'e algebra, including the Hamiltonian,
are the constraint equations of the gravitational field.
They can be written as
\begin{gather}
	\frac{1}{16\pi\sqrt{\gamma}} \left[ \gamma \text{R}
		+ \frac{1}{2} \left( \gamma_{ij} \pi^{ij} \right)^2
		- \gamma_{ij} \gamma_{k l} \pi^{ik} \pi^{jl}\right]
		= \mathcal{H}^{\text{matter}} \,, \label{ham} \\
	- \frac{1}{8\pi} \gamma_{ij} \pi^{jk}_{~~ ; k} = \mathcal{H}^{\text{matter}}_i \,, \label{mom}
\end{gather}
with the definitions
\begin{align}
\pi^{ij} &= - \sqrt{\gamma} (\gamma^{ik}\gamma^{jl} - \gamma^{ij}\gamma^{kl})K_{kl} \,, \\
\mathcal{H}^{\rm matter} &= \sqrt{\gamma}T_{\mu\nu} n^{\mu}n^{\nu} \,, \\
\mathcal{H}^{\rm matter}_i &= - \sqrt{\gamma}T_{i \nu} n^{\nu} \,,
\end{align}
and arise as certain projections of the Einstein equations with respect to a timelike
unit 4-vector $n_{\mu}$ with components $n_{\mu} = (-N, 0, 0, 0)$ or $n^{\mu} = (1, -N^i) / N$.
Here $\gamma_{ij}$ is the induced three-dimensional metric of the hypersurfaces orthogonal to $n_{\mu}$,
$\gamma$ its determinant, $\text{R}$ the three-dimensional Ricci scalar, $K_{ij}$ the extrinsic curvature,
$N$ the lapse function, $N^i$ the shift vector, $ \sqrt{\gamma}T_{\mu\nu}$ the stress-energy
tensor density of the matter system, and $;$ denotes the three-dimensional covariant derivative.
Partial derivatives are indicated by a comma.

For nonspinning objects $\frac{1}{16\pi} \pi^{ij}$ is the canonical momentum
conjugate to $\gamma_{ij}$ before gauge fixing. For spinning
objects we now make an ansatz for the canonical field momentum
of the form
\begin{align}\label{pican}
	\pi^{ij}_{\text{can}} &= \pi^{ij} + \pi^{ij}_{\text{matter}} \,,
\end{align}
where $\pi^{ij}_{\text{matter}}$ shall be linear in the spins
and will be fixed later on.
% We will only consider the case $\pi^{ii}_{\text{matter}} = 0$ here.
In the ADM transverse-traceless (ADMTT) gauge defined by
\begin{subequations}\label{ADMTT}
\begin{align}
3\gamma_{ij,j} - \gamma_{jj,i} &= 0 \,, \label{ADMTTg} \\
\pi^{ii}_{\text{can}} &= 0 \,, \label{ADMTTpi}
\end{align}
\end{subequations}
which will be used throughout this paper, one has the decompositions
\begin{align}
	\gamma_{ij} &= \left( 1 + \frac{\phi}{8} \right)^4 \delta_{ij} + h^{\text{TT}}_{ij} \,,
		\label{gdecomp} \\
	\pi^{ij}_{\text{can}} &= \pi^{ij\text{TT}}_{\text{can}} + \tilde{\pi}^{ij}_{\text{can}} \,,
\end{align}
where $h^{\text{TT}}_{ij}$ and $\pi^{ij\text{TT}}_{\text{can}}$ are transverse-traceless,
e.g, $h^{\text{TT}}_{ii} = h^{\text{TT}}_{ij,j}=0$, and $\tilde{\pi}^{ij}_{\text{can}}$ is
related to vector potentials $V^i_{\text{can}}$ and $\tilde{\pi}^i_{\text{can}}$ by
\begin{align}
\tilde{\pi}^{ij}_{\text{can}} &= V^i_{\text{can}, j}
	+ V^j_{\text{can}, i} - \frac{2}{3} \delta_{ij} V^k_{\text{can}, k} \,, \\
	&= \tilde{\pi}^i_{\text{can}, j} + \tilde{\pi}^j_{\text{can}, i}
	- \frac{1}{2} \delta_{ij} \tilde{\pi}^k_{\text{can}, k}
	- \frac{1}{2} \Delta^{-1} \tilde{\pi}^k_{\text{can}, ijk} \,.
\end{align}
It holds
\begin{align}
V^i_{\text{can}} &= \left( \delta_{ij} - \frac{1}{4} \partial_i \partial_j \Delta^{-1} \right) \tilde{\pi}^j_{\text{can}} \,, \\ \tilde{\pi}^i_{\text{can}} &= \Delta^{-1} \pi^{ij}_{\text{can},j}
	= \Delta^{-1} \tilde{\pi}^{ij}_{\text{can},j} \,, \\
\pi^{ij\text{TT}}_{\text{can}} &= \delta^{\text{TT}ij}_{kl} \pi^{kl}_{\text{can}} \,,
\end{align}
with the inverse Laplacian $\Delta^{-1}$, the partial space-coordinate derivatives $\pa_i$ and
\begin{equation}\label{TTproj}
\begin{split}
\delta^{\text{TT}kl}_{ij} &= \tfrac{1}{2} [(\delta_{il}-\Delta^{-1}\pa_{i}\pa_{l})(\delta_{jk}-\Delta^{-1}\pa_{j}\pa_{k}) \nl
+(\delta_{ik}-\Delta^{-1}\pa_{i}\pa_{k})(\delta_{jl}-\Delta^{-1}\pa_{j}\pa_{l}) \nl -(\delta_{kl}-\Delta^{-1}\pa_{k}\pa_{l})(\delta_{ij}-\Delta^{-1}\pa_{i}\pa_{j})] \,.
\end{split}
\end{equation}
See Sec.\ \ref{SPN} for details on the decompositions for $\gamma_{ij}$ and $\pi^{ij}_{\text{can}}$.
Notice that the form of the trace term in (\ref{gdecomp})
is adapted to the Schwarzschild metric in isotropic coordinates,
with obvious advantages for perturbative expansions.

% The same decomposition is given for $\pi^{ij}_{\text{can}}$ in
% terms of $\tilde{\pi}^{ij}_{\text{can}}$, $\pi^{ij\text{TT}}_{\text{can}}$,
% $V^i_{\text{can}}$ and $\tilde{\pi}^i_{\text{can}}$.
Now the four field constraints can be solved for the four variables $\phi$ and $\tilde{\pi}^i_{\text{can}}$
in terms of $h^{\text{TT}}_{ij}$, $\pi^{ij\text{TT}}_{\text{can}}$
and matter variables, which enter through the source terms
$\mathcal{H}^{\rm matter}$ and $\mathcal{H}^{\rm matter}_i$.
An analytic solution for $\phi$ and $\tilde{\pi}^i_{\text{can}}$, however,
can in general only be given in some approximation scheme.

In the ADMTT gauge the momentum constraint (\ref{mom}) can \emph{exactly} be written as
\begin{equation}
\begin{split}\label{cmom}
\tilde{\pi}^{ij}_{\text{can},j} &=
	- 8\pi ( \mathcal{H}^{\text{matter}}_{i}
		+ \mathcal{H}^{\pi \text{matter}}_i ) \nl
	+ A^{ij}_{\text{can},j}
	- \Delta \left( V^k_{\text{can}} h_{ki}^{\text{TT}} \right) \nl
	+ \frac{1}{2} \pi^{jk\text{TT}}_{\text{can}} h_{jk,i}^{\text{TT}}
	- ( \pi^{jk\text{TT}}_{\text{can}} h_{ki}^{\text{TT}} )_{,j} \,,
\end{split}
\end{equation}
with the definitions
\begin{align}
\begin{split}
A^{ij}_{\text{can}} &= \left[ 1 - \left( 1 + \tfrac{1}{8} \phi \right)^4 \right] ( \tilde{\pi}^{ij}_{\text{can}} + \pi^{ij\text{TT}}_{\text{can}} ) \nl
	+ V^k_{\text{can}} ( h_{ki,j}^{\text{TT}} + h_{kj,i}^{\text{TT}} - h_{ij,k}^{\text{TT}} )
	- \frac{1}{3} V^k_{\text{can},k} h_{ij}^{\text{TT}} \,,
\end{split}\\
\mathcal{H}^{\pi \text{matter}}_i &= \frac{1}{16\pi} [
	- 2 ( \gamma_{ik} \pi^{kj}_{\text{matter}} )_{,j}
	+ \pi^{jk}_{\text{matter}} \gamma_{jk,i} ] \,.
\end{align}
This equation will allow us to derive explicit expressions for total linear
and angular momentum without solving the constraints. Notice that
$A^{ij}_{\text{can}} = A^{ji}_{\text{can}}$ and $A^{ii}_{\text{can}} = 0$. \\

\subsection{Global Poincar\'e algebra}\label{pgen}
The global Poincar\'e algebra is a consequence of the
asymptotic flatness and is represented by Poisson brackets
of the corresponding conserved quantities. These quantities
are the ADM energy $E$, total linear momentum $P_i$,
total angular momentum $J_i = \frac{1}{2} \epsilon_{ijk} J_{jk}$,
and the boost vector $K^i$. They are given by surface integrals at
spatial infinity. The boosts have an explicit dependence on the time
$t$ and can be decomposed as $K^i = G^i - t P_i$, where
$X^i = G^i / E$ is the coordinate of the center-of-mass.
$G^i$ will be called center-of-mass vector in the following.
The corresponding surface integrals read, with spatial coordinates denoted $x^i$,
\begin{align}
	E &= \frac{1}{16\pi}\oint d^2 s_i (\gamma_{ij,j} -\gamma_{jj,i}) \,, \\
	G^i &= \frac{1}{16\pi}\oint d^2 s_k \left[ x^i ( \gamma_{kl,l} - \gamma_{ll,k} )
		- \gamma_{ik} + \delta_{ik} \gamma_{ll} \right]\,, \\
	P_i &= - \frac{1}{8\pi}\oint d^2 s_k \pi^{ik} \,, \\
	J_{ij} &= - \frac{1}{8\pi}\oint d^2 s_k ( x^i \pi^{jk} - x^j \pi^{ik}) \,.
\end{align}
See, e.g., \cite{Regge:Teitelboim:1974}. Using the gauge conditions and also the momentum constraint
in the form (\ref{cmom}), these surface
integrals can be transformed into the volume integrals
\begin{align}
	E &= - \frac{1}{16\pi} \int d^3x \, \Delta \phi \,, \\
	G^i &= - \frac{1}{16\pi} \int d^3x \, x^i \Delta \phi \,, \label{Gint} \\
	P_i &= P_i^{\text{matter}}
		- \frac{1}{16\pi} \int d^3x \, \pi_{\text{can}}^{kl\text{TT}} h^{\text{TT}}_{kl,i} \,, \label{Ptot} \\
\begin{split}
	J_{ij} &= J_{ij}^{\text{matter}} - \frac{1}{16\pi} \int d^3x \, 2 ( \pi_{\text{can}}^{ik\text{TT}} h^{\text{TT}}_{kj}
			- \pi_{\text{can}}^{jk\text{TT}} h^{\text{TT}}_{ki} ) \nl
		- \frac{1}{16\pi} \int d^3x \, ( x^i \pi_{\text{can}}^{kl\text{TT}} h^{\text{TT}}_{kl,j} - x^j \pi_{\text{can}}^{kl\text{TT}} h^{\text{TT}}_{kl,i} ) \label{Jtot} \,,
\end{split}
\end{align}
with the matter parts
\begin{align}
	P_i^{\text{matter}} &= \int d^3x \, ( \mathcal{H}^{\rm matter}_i + \mathcal{H}^{\pi \text{matter}}_i ) \,, \\
\begin{split}
J_{ij}^{\text{matter}} &= \int d^3x \, ( x^i \mathcal{H}^{\rm matter}_j
		+ x^i \mathcal{H}^{\pi \rm matter}_j \nlq
		- x^j \mathcal{H}^{\rm matter}_i - x^j \mathcal{H}^{\pi \rm matter}_i ) \,.
\end{split}
\end{align}
Here we used the fact that $\pi^{ij}_{\text{matter}}$ has a compact support.

Now we require that the matter parts of total linear and angular momentum are
of the form
\begin{align}
	P_i^{\text{matter}} &= \sum_a P_{ai} \,, \label{Pstd} \\
	J_{ij}^{\text{matter}} &= \sum_a ( \hat{z}_a^i P_{aj} - \hat{z}_a^j P_{ai} ) + \sum_a S_{a(i)(j)} \,, \label{Jstd}
\end{align}
where $\hat{z}_a^i$, $P_{aj}$, and $S_{a(i)(j)} = \epsilon_{ijk} S_{a(k)}$
are the canonical position, momentum, and spin of the particles,
because this is the expected form for standard canonical variables
with equal-time Poisson brackets
\begin{align}
\{ h^{\text{TT}}_{ij}({\bf x}), \pi^{kl\text{TT}}_{\text{can}}({\bf x}') \}
	&= 16\pi \delta^{\text{TT}kl}_{ij}\delta({\bf x} - {\bf x}') \,, \\
\{ \hat{z}^i_a, P_{a j} \} &= \delta_{ij} \,, \\
\{ S_{a(i)}, S_{a(j)} \} &= \epsilon_{ijk} S_{a(k)} \,,
\end{align}
zero otherwise, where $\epsilon_{ijk} = \frac{1}{2} (i-j)(j-k)(k-i)$.
Equations (\ref{Pstd}) and (\ref{Jstd}) ensure that a great part of the Poincar\'e algebra
is fulfilled; see Appendix \ref{poincare}. With the definition
\begin{equation}
	\pi^{ij}_{\text{matter}} = 16 \pi \sum_a \pi^{ij}_a \delta_a \,,
\end{equation}
where $\delta_a = \delta(\vct{x} - \hat{\vct{z}}_a)$
with normalization $\int d^3x \, \delta_a = 1$,
the source of the momentum constraint $\mathcal{H}^{\rm matter}_{i}$
then is of the form\footnote{Here we assumed
that the variables from different objects do not mix
(e.g., as in $P_1 \delta_2$) at this stage.}
\begin{align}
	\mathcal{H}^{\rm matter}_{i} &= \sum_a \bigg[ (P_{ai} - \pi^{jk}_a \gamma_{jk,i}) \delta_a
		+ \frac{1}{2} (s_a^{ij} \delta_a)_{,j} \bigg] \,, \label{defP} \\
	S_{a(i)(j)} &= s_a^{[ij]}
		+ 2 \pi_a^{ik} h^{\text{TT}}_{kj}
		- 2 \pi_a^{jk} h^{\text{TT}}_{ki} \,. \label{Scons}
\end{align}
At linear order in spin and 3.5PN, the first equation defines the canonical momentum,
while the second one fixes the canonical position, the triad (i.e., the local basis of the canonical spin) and $\pi_a^{ij}$, up to canonical transformation. Further the Euclidean spin length $s_a$ given by
\begin{equation}\label{spinSq}
	2 s_a^2 = 2 S_{a(i)} S_{a(i)} = S_{a(i)(j)} S_{a(i)(j)} \,,
\end{equation}
has vanishing Poisson bracket with all
quantities, including the Hamiltonian. Thus $s_a$ must be a constant of motion.

The ADM Hamiltonian $H_{\text{ADM}}$ results as
\begin{equation}\label{HADM}
H_{\text{ADM}} = - \frac{1}{16\pi} \int d^3x \, \Delta
	\phi[ \hat{z}^i_a, P_{a i}, S_{a(i)}, h^{\text{TT}}_{ij}, \pi_{\text{can}}^{ij\text{TT}} ] \,.
\end{equation}
This is the ADM energy depending on the canonical variables. It
arises from solving the constraints for $\phi$, once the source terms
of the constraints, $\mathcal{H}^{\text{matter}}$ and $\mathcal{H}^{\text{matter}}_{i}$,
are expressed in terms of the canonical variables.
An action corresponding to $H_{\text{ADM}}$ is given by (4.33) in \cite{steinhoff08:spin2}
or (51) in \cite{SSepl}.

Total linear and angular momentum could, of course, also be represented on the
phase space in a more complicated way than given by (\ref{Ptot}), (\ref{Jtot}),
(\ref{Pstd}), and (\ref{Jstd}). However, the ADMTT gauge manifestly respects
the Euclidean group in its standard representation, which implies that its generators
$P_i$ and $J_{ij}$ are also in its standard representation on the phase space;
see Appendix \ref{poincare} and also \cite{DJS00}.

\section{The source\label{Ssource}}
\subsection{(3+1)-split}
The stress-energy tensor density to linear order in spin is given by
\cite{Tulczyjew:1959,Dixon:1979,T02}
\begin{equation}
\sqrt{-g} T^{\mu\nu} = \sum_a \int d \tau \bigg[
	m_a u^{\mu}_a u^{\nu}_a \delta_{(4)a}
	+ ( u^{(\mu}_a S^{\nu)\alpha}_a \delta_{(4)a} )_{||\alpha}
\bigg] \,,
\end{equation}
in the covariant spin supplementary condition (SSC)
\begin{equation}
S^{\mu\nu}_a u_{a\nu} = 0 \,.
\end{equation}
Here $m_a$ is the mass, $u_a$ the 4-velocity, $\tau$ the
proper time parameter, $S^{\mu\nu}_a$ the spin tensor,
$||$ denotes the four-dimensional covariant derivative,
and $\delta_{(4)a} = \delta(x - z_a)$
with normalization $\int d^4x \, \delta_{(4)a} = 1$.
$z_a^{\mu}$ is the coordinate of the $a$-th object.
The matter EOM, i.e., the Mathisson-Papapetrou equations
\cite{Mathisson:1937,papapetrou51a,T02},
in covariant SSC and at linear order in spin
can be followed from $T^{\mu\nu}_{~~||\nu}=0$ as
\begin{subequations}\label{MPeqs}
\begin{align}
	\frac{D S_a^{\mu\nu}}{D \tau} &= 0 \,, \label{spinEOM} \\
	\frac{D p_a^{\mu}}{D \tau} &= - \frac{1}{2} {}^{(4)}\text{R}^{\mu}_{~\rho\beta\alpha}
		u^{\rho}_a S^{\beta\alpha}_a \,, \\
	\frac{d z_a^\mu}{d \tau} &\equiv u^{\mu}_a = \frac{p_a^{\mu}}{m_a} \,.
\end{align}
\end{subequations}
Here ${}^{(4)}\text{R}^{\mu}_{~\rho\beta\alpha}$ is the four-dimensional Riemann
tensor and $D$ the four-dimensional covariant parameter derivative. The spin length
$s_a$ is given by $2 s_a^2 = S_a^{\mu\nu} S_{a\mu\nu}$ and
obviously is a constant of motion due to (\ref{spinEOM}).

The (3+1)-split of $u_a^2 = -1$, the SSC and the spin length reads
\begin{align}
	np_a &= n^{\mu} p_{a \mu} = - \sqrt{m_a^2 + \gamma^{ij} p_{a i} p_{a j}} \,, \\
	nS_{ai} &= n^{\mu} S_{a \mu i} = \frac{p_{ak} \gamma^{kj} S_{aji}}{np_a} \,, \\
	2 s_a^2 &= \gamma^{ki} \gamma^{lj} S_{a kl} S_{a ij}
		- 2 nS_{ai} nS_a^i \,. \label{covlength}
\end{align}
Notice that $nS_a^i = \gamma^{ij} nS_{aj}$.
The components of the stress-energy tensor density
are given by, with $\delta_a = \delta(\vct{x}-\vct{z}_a)$,
\begin{equation}
\mathcal{H}^{\text{matter}} = \sum_a \Bigg[ - np_a \delta_a
	- K^{kl} \frac{p_{ak} nS_{al}}{np_a} \delta_a - ( nS_a^k \delta_a )_{;k} \Bigg] \,,
\end{equation}
\begin{widetext}
\begin{align}
\mathcal{H}^{\text{matter}}_i &= \sum_a \Bigg[ p_{ai} \delta_a + K_{ij} nS_a^j \delta_a
	+ \bigg( \frac{1}{2} \gamma^{mk} S_{aik} \delta_a
		+ \delta_i^{(k} \gamma^{l)m} \frac{p_{ak} nS_{al}}{np_a} \delta_a \bigg)_{;m} \Bigg] \,, \\
\begin{split}
\mathcal{T}_{i j} &= \sum_a \Bigg[ \bigg( - \frac{p_{a i} p_{a j}}{np_a}
		+ S_{a k (i} K^k_{j)} + \frac{p_{a (i} S_{a j) k} p_{a l} K^{kl}}{(np_a)^2}
		- \frac{nS_{a k} p_{a (i} K_{j)}^k}{np_a}
		-  \frac{p_{a k} nS_{a (i} K_{j)}^k}{np_a}
		+ \frac{p_{a (i} nS_{a j)} p_{a k} p_{a l} K^{kl}}{(np_a)^3}
	\bigg) \delta_a \nlq
	+ \bigg( \gamma^{kl} \frac{S_{a l ( i} p_{a j )}}{np_a} \delta_a
	- \gamma^{kl} \frac{p_{a l} p_{a (i} nS_{a j)}}{(np_a)^2} \delta_a \bigg)_{;k} \Bigg] \,, \label{Tij}
\end{split}
\end{align}
where $\mathcal{T}_{i j} = \sqrt{\gamma} T_{i j}$.
After transition to Newton-Wigner (NW) variables
\begin{subequations}\label{NWvars}
\begin{align}
	S_{a ij} &= \hat{S}_{a ij} - \frac{p_{a i} nS_{a j}}{m_a-np_a} + \frac{p_{a j} nS_{a i}}{m_a-np_a} \,, \qquad
	nS_{a i} = - \frac{p_{a k} \gamma^{kj} \hat{S}_{a ji}}{m_a} \,, \label{NWspin} \\
	z^i_a &= \hat{z}^i_a - \frac{nS^i_a}{m_a - np_a} + \delta z^i_a \,, \label{NWpos} \\
	p_{ai} &= P_{ai} - nS^k_a K_{ik} - \pi^{jk}_{a} \gamma_{jk,i}
		+ \left[ \frac{1}{2} \gamma^{kl}\Gamma^j_{li}
		- \frac{P_{am} P_{aq} \gamma^{mj}}{nP_a(m_a-nP_a)} \gamma^{l(q} \Gamma^{k)}_{li}
		+ \frac{P_{ap} P_{aq} \gamma^{qj} \gamma^{km}}{m_a (m_a-nP_a)} \Gamma^p_{mi} \right] \hat{S}_{ajk} \,, \label{NWmom}
\end{align}
\end{subequations}
where $\Gamma^{k}_{ij}$ are the three-dimensional Christoffel symbols,
the source expressions of the constraints
read [now $\delta_a = \delta(\vct{x}-\hat{\vct{z}}_a)$]
\begin{align}
\begin{split}
\mathcal{H}^{\text{matter}} &= \sum_a \Bigg[ - nP_a \delta_a
	- \frac{1}{2} \bigg( \frac{\hat{S}_{ali} P_{aj}}{nP_a}
		+ \gamma^{mn}\frac{\hat{S}_{ami} P_{aj} P_{an} P_{al}}{(nP_a)^2(m_a-nP_a)}
		+ 2 \frac{P_{al} \pi_{aij}}{nP_a}
		+ \frac{P_{ai} P_{aj}}{nP_a} \delta z_{al}
	\bigg) \gamma^{kl} \gamma^{ij}_{~~,k} \delta_a \nlq
	- \bigg( \frac{P_{al}}{m_a-nP_a}\gamma^{ij}\gamma^{kl} \hat{S}_{ajk} \delta_a
		- nP_a \delta z^i_a \delta_a \bigg)_{,i} \Bigg] \label{NWcham} \,,
\end{split}\\
\mathcal{H}^{\text{matter}}_i &= \sum_a \Bigg[ P_{ai}\delta_a - \pi^{jk}_{a} \gamma_{jk,i} \delta_a
	+ \frac{1}{2} \bigg(\gamma^{mk}\hat{S}_{aik} \delta_a -\frac{P_{al}P_{ak}}{nP_a (m_a-nP_a)}
		(\gamma^{mk}\delta_i^p + \gamma^{mp} \delta_i^k)\gamma^{ql} \hat{S}_{aqp}\delta_a
		- 2 P_{ai} \delta z^m_a \delta_a \bigg)_{,m} \Bigg] \label{NWcmom} \,.
\end{align}
\end{widetext}
Obviously (\ref{NWcmom}) is now of the from (\ref{defP}), which uniquely
fixed the relation between covariant linear momentum $p_{ai}$
and canonical momentum $P_{ai}$, Eq.\ (\ref{NWmom}).
The spin redefinition (\ref{NWspin}) transforms the spin length
(\ref{covlength}) into
\begin{equation}
	2 s_a^2 = \gamma^{ki} \gamma^{lj} \hat{S}_{a kl} \hat{S}_{a ij} = \hat{S}_{a(i)(j)} \hat{S}_{a(i)(j)} \,,
\end{equation}
where $\hat{S}_{a(i)(j)}$ are the components of $\hat{S}_{a ij}$ in some
local Euclidean basis. Comparing with (\ref{spinSq}) suggests that
$\hat{S}_{a(i)(j)}$ is equal to the canonical spin $S_{a(i)(j)}$.
However, a local Euclidean basis is only unique up to a rotation.
Fortunately Eq.\ (\ref{Scons}) will be seen to
uniquely fix a basis such that $\hat{S}_{a(i)(j)}$ and $S_{a(i)(j)}$ can
be identified in that basis. (Notice that $S_{a(i)(j)}$ are \emph{not}
the components of the covariant spin $S_{a ij}$ in a local basis here,
in contrast to \cite{SSepl}.)
The redefinition of the position (\ref{NWpos}) consists of a term known
from flat-space and a yet unknown quantity $\delta z^i_a$, which
will also be fixed by Eq.\ (\ref{Scons}) later on.

\subsection{Triad}
The relation between $\hat{S}_{a ij}$ and $\hat{S}_{a(i)(j)}$ can be written
with the help of a triad $e^{i(j)}$ as
\begin{equation}\label{espin}
	\hat{S}_{a(i)(j)} = e^{k(i)} e^{l(j)} \hat{S}_{akl} \,.
\end{equation}
This spin has a constant Euclidean length for all choices of $e^{k(i)}$.
Notice that the triad is needed only on the worldlines and not
as a field over the entire spacetime here.
The triad can be split into symmetric $\tilde{e}^{i(j)}$ and antisymmetric $\hat{e}^{i(j)}$ parts
as $e^{i(j)} = \tilde{e}^{i(j)} + \hat{e}^{i(j)}$.
Perturbative expansion of $e^{i(k)} e^{j(k)} = \gamma^{ij}$
leads to\footnote{Notice that in this formula the \emph{subscripts}
in round brackets denote the formal order in $c^{-1}$, not an
index in a local basis.}
\begin{equation}
	e^{i(j)}_{(n)} = \frac{1}{2} \gamma^{ij}_{(n)} - \frac{1}{2} \sum_{m=1}^{n-1} e^{i(k)}_{(m)} e^{j(k)}_{(n-m)}
		+ \hat{e}^{i(j)}_{(n)} \,,
\end{equation}
where $\gamma^{ij}_{(0)}=\delta_{ij}$ and $e^{i(k)}_{(0)}=\delta_{ik}$
was assumed. For example, the leading order results are:
\begin{align}
	e^{i(j)}_{(2)} &= \hat{e}^{i(j)}_{(2)} - \frac{1}{4} \delta_{ij} \phi_{(2)} \\
	e^{i(j)}_{(4)} &= \hat{e}^{i(j)}_{(4)} - \frac{1}{2} \hat{e}^{i(k)}_{(2)} \hat{e}^{j(k)}_{(2)}
		- \frac{1}{4} \delta_{ij} \phi_{(4)} + \frac{3}{64} \delta_{ij} \phi_{(2)}^2 - \frac{1}{2} h^{\text{TT}}_{ij}
\end{align}
The symmetric part is thus fixed. The antisymmetric part $\hat{e}^{i(j)}$, however,
must be imposed, as it represents the three rotational degrees of freedom
left in the definition of the local basis.
Thus $\hat{e}^{i(j)}$ represents the degrees of freedom left in the definition
of the canonical spin variable.

\subsection{Fixation of the NW variables}
Whereas the canonical momentum was already unambiguously fixed by (\ref{defP}) as (\ref{NWmom}),
$\delta z^i_a$, $\hat{e}^{i(j)}$, and $\pi_a^{ij}$ are still unknown.
We will see now that these can be fixed up to a canonical transformation
with the help of (\ref{Scons}).
For our source one gets for the leading orders of $s_{a}^{[ij]}$
\begin{align}
s_{a(3)}^{ij} &= S_{a(i)(j)} \,, \\
\begin{split}
s_{a(5)}^{[ij]} &= \hat{e}^{i(k)}_{(2)} S_{a(k)(j)} - \hat{e}^{j(k)}_{(2)} S_{a(k)(i)} \nl
	- P_{ai} \delta z^j_{a(2)} + P_{aj} \delta z^i_{a(2)} \,,
\end{split}
\end{align}
and thus from (\ref{Scons}) one concludes
\begin{equation}
	\hat{e}^{i(k)}_{(2)} = 0 \,, \quad
	\delta z^i_{a(2)} = 0 \,.
\end{equation}
It is crucial that $\hat{e}^{i(k)}_{(2)}$ must be antisymmetric.
At the next order it holds
\begin{equation}
\begin{split}
s_{a(7)}^{[ij]} &= \hat{e}^{i(k)}_{(4)} S_{a(k)(j)} - \hat{e}^{j(k)}_{(4)} S_{a(k)(i)} \nl
	- P_{ai} \delta z^j_{a(4)} + P_{aj} \delta z^i_{a(4)} \,.
\end{split}
\end{equation}
Further $\pi_{a(3)}^{ij} = 0$ because $\pi_a^{ij}$ is linear in spin, symmetric, and
contains an even number of momentum variables (because of parity).
Equation (\ref{Scons}) then leads to
\begin{equation}
	\hat{e}^{i(k)}_{(4)} = 0 \,, \quad
	\delta z^i_{a(4)} = 0 \,.
\end{equation}

For $s^{[ij]}_{a (9)}$ one has
\begin{equation}
\begin{split}
	s^{[ij]}_{a (9)} &= \hat{e}^{i(k)}_{(6)} S_{a(k)(j)} - \hat{e}^{j(k)}_{(6)} S_{a(k)(i)} \nl
		- P_{ai} \delta z^j_{a(6)} + P_{aj} \delta z^i_{a(6)} \nl
		- \frac{1}{2m_a^2} P_{a k} S_{a (k) (l)} P_{a [i} h_{j] l}^{\text{TT}} \nl
		- \frac{1}{2m_a^2} P_{a k} P_{a l} S_{a k [i} h_{j] l}^{\text{TT}} \,.
\end{split}
\end{equation}
The most general (sensible) solution of (\ref{Scons}) at this order is
\begin{subequations}\label{Cambig}
\begin{align}
\pi^{ij}_{(5)a} &= \frac{1-C}{8m_a^2} ( P_{a i} P_{a k} S_{a (k) (j)} + P_{a j} P_{a k} S_{a (k) (i)} ) \,, \\
\hat{e}^{i(j)}_{(6)} &= \frac{C}{2m_a^2} P_{ak} P_{a[i} h_{j] k}^{\text{TT}} \,, \\
\delta z^i_{a(6)} &= \frac{C}{4m_a^2} P_{aj} ( S_{a(k)(i)} h_{jk}^{\text{TT}} + S_{a(k)(j)} h_{ik}^{\text{TT}} ) \,,
\end{align}
\end{subequations}
with an arbitrary constant $C$. Notice that $\pi^{ii}_{(5)a} = 0$.
Now we can remove the ambiguity $C$ by a canonical transformation generated by
\begin{equation}
	g = \frac{C}{4m_a^2} P_{a i} P_{a k} S_{a (k) (j)} \int d^3x \, h^{\text{TT}}_{i j} \delta_a \,,
\end{equation}
which transforms an arbitrary phase space function $A$ as
\begin{equation}
	A \rightarrow A + \{ A , g \} \,,
\end{equation}
to the required order. For our fundamental variables this means
\begin{align}
	h^{\text{TT}}_{i j} &\rightarrow h^{\text{TT}}_{i j} \,, \\
\begin{split}
	\pi^{i j \text{TT}}_{\text{can}} &\rightarrow
		\pi^{i j \text{TT}}_{\text{can}}
			- \delta^{\text{TT}ij}_{kl} \sum_a \frac{4\pi C}{m_a^2} P_{a k} P_{a m} S_{a (m) (l)} \delta_a \,,
\end{split}\\
	S_{a(i)(j)} &\rightarrow S_{a(i)(j)} - \hat{e}^{i(k)}_{(6)} S_{a(k)(j)} - \hat{e}^{j(k)}_{(6)} S_{a(i)(k)} \,, \\
	\hat{z}^i_a &\rightarrow \hat{z}^i_a - \delta z^i_{a(6)} \,, \\
	P_{a i} &\rightarrow P_{a i} - \frac{C}{4m_a^2} P_{a l} P_{a j} S_{a (j) (k)} h^{\text{TT}}_{kl,i} \,.
\end{align}
This indeed removes all terms depending on $C$ from the source expressions
in (\ref{NWcham}) and (\ref{NWcmom}) at the considered order. We can therefore choose $C=0$.
This choice has the nice properties that
the triad fulfills the gauge condition $e^{i(j)} = e^{j(i)}$ and
that the transition to the NW position $\hat{z}^i_a$ can be expressed in terms of a
Lie-shift; see Appendix \ref{lie}. Further this choice agrees with the
action approach in \cite{SSepl}.

\subsection{Comparison with the action approach}
For the action approach in \cite{SSepl} the gauge condition
$e_{i(j)} = e_{j(i)} = e_{ij}$ holds to all orders,
which fixes $e_{ij}$ as the matrix square-root of the
three-dimensional metric $e_{ij} e_{jk} = \gamma_{ik}$, or
\begin{equation}\label{groot}
e_{ij} = \sqrt{(\gamma_{kl})} \,.
\end{equation}
Further, the transition to NW variables in \cite{SSepl} agrees with (\ref{NWvars})
for $\delta z^i_a = 0$ to all orders. Notice that
Eq. (31) in \cite{SSepl} is written in terms of the
covariant spin and one has to use Eq.\ (45) in \cite{SSepl} and (\ref{defpia}) on the triad terms.
It is shown in \cite{SSepl} that a suitable choice of $\pi_a^{ij}$
extends the ADM formalism for spinning objects to all PN orders
linear in spin. According to \cite{SSepl}, it holds
\begin{align}\label{defpia}
\pi_a^{ij} &= \frac{1}{2} \hat{A}^{(ij)}_a + B^{ij}_{kl} \hat{A}^{[kl]}_a \,, \\
\gamma_{ik} \gamma_{jl} \hat{A}^{kl}_a &= \frac{1}{2} \hat{S}_{aij} + \frac{m_a P_{a(i} nS_{aj)}}{nP_a (m_a-nP_a)} \,,
\end{align}
where the quantity $B^{kl}_{ij}$ is defined by
\begin{equation}
e_{k[i} e_{j]k,\mu} = B^{kl}_{ij} \gamma_{kl,\mu} \,. \label{eliminatee}
\end{equation}
This can also be written as
\begin{equation}\label{Bder}
2 B^{kl}_{ij} = e_{mi} \frac{\pa e_{mj}}{\pa \gamma_{kl}} - e_{mj} \frac{\pa e_{mi}}{\pa \gamma_{kl}} \,,
\end{equation}
which must be evaluated perturbatively using Eq.\ (4.22) in
\cite{steinhoff08:spin2}, e.g.,
\begin{equation}
2 B^{kl}_{ij} = \frac{1}{4} ( \delta_{j ( k} h^{\text{TT}}_{l ) i}
	- \delta_{i ( k} h^{\text{TT}}_{l ) j} ) + \Order((h^{\text{TT}})^2) \,.
\end{equation}
Further it holds
\begin{align}\label{tacepia}
\pi_a^{ii} = \delta_{kl} \gamma^{ki} \gamma^{lj} \frac{m_a P_{ai} nS_{aj}}{2 nP_a (m_a-nP_a)} \,,
\end{align}
which follows from $B^{kl}_{ij} \delta_{kl} = 0$. This leads to a deviation
of our gauge condition $\pi^{ii}_{\text{can}} = 0$ from the original
ADM one $\pi^{ii} = 0$ at the formal 5PN order.

We will now show that (\ref{Scons}) exactly holds
for the action approach. This gives a check of the action
approach to all orders, for any approximation scheme.
From the discussion above it is clear that
the source of the momentum constraint is of the form (\ref{defP}).
Further it holds $s_a^{ij} = \gamma_{ik} \hat{A}^{kj}_a$.
Equation (\ref{Scons}) then leads to the condition
\begin{equation}
\begin{split}
0&=S_{a(k)(l)} ( e_{ki} e^{lj} - e_{kj} e^{li} - 2 \delta_{ki} \delta_{lj} \nl
	- 2 e^{km} e^{ln} B_{mn}^{jp} \gamma_{pi} + 2 e^{km} e^{ln} B_{mn}^{ip} \gamma_{pj} ) \,, \label{actioncheck}
\end{split}
\end{equation}
Evaluating
\begin{equation}
\frac{\pa e_{mn}}{\pa \gamma_{pq}} \frac{\pa (e_{pk} e_{kq})}{\pa e_{ij}} = \frac{\pa e_{mn}}{\pa e_{ij}} \,,
\end{equation}
results in
\begin{equation}
\frac{\partial e_{mn}}{\partial \gamma_{jp}} \gamma_{pi} = e_{i(m} \delta_{n)j} - \frac{\partial e_{mn}}{\partial \gamma_{pq}} e_{pi} e_{qj} \,.
\end{equation}
The second term on the right-hand side is symmetric in $i$ and $j$ and cancels
from (\ref{actioncheck}) with (\ref{Bder}) inserted.
It is then easy to see that (\ref{actioncheck}) and
thus (\ref{Scons}) are fulfilled.

The action approach shows that (\ref{NWcham}) and (\ref{NWcmom})
can be applied to all orders, if $\delta z^i_a = 0$, $\hat{e}^{i(j)} = 0$,
and (\ref{defpia}) are inserted.
Further, if the spin correction to the field momentum
$\pi^{ij}_a$ is neglected, (\ref{NWcham}) and (\ref{NWcmom})
coincide with (4.23) and (4.25) in \cite{steinhoff08:spin2}.
In \cite{steinhoff08:spin2} a Lie-shift was used to redefine
the position instead of a Taylor expansion here; see Appendix \ref{lie}.

\section{PN expansion\label{SPN}}
In this section we give the PN expansion of the field
constraints relevant for the ADM Hamiltonian up to
and including 3.5PN. Here and in Sec.\ \ref{Sflux}
we made use of xTensor \cite{M02} (a free package
for Mathematica \cite{W03}), especially of its fast index
canonicalizer based on the package xPerm \cite{M08}.

For $\pi^{ij}$ we use the general decomposition
\begin{equation}\label{decomp}
\pi^{ij} = \pi^{ij\text{TT}} + \tilde{\pi}^{ij} + \hat{\pi}^{ij} \,,
\end{equation}
with
\begin{subequations}\label{decompparts}
\begin{align}
\pi^{ij\text{TT}} &= \delta^{\text{TT}ij}_{kl} \pi^{kl} \,, \\
\tilde{\pi}^{ij} &= \tilde{\pi}^i_{, j} + \tilde{\pi}^j_{, i}
	- \frac{1}{2} \delta_{ij} \tilde{\pi}^k_{, k}
	- \frac{1}{2} \Delta^{-1} \tilde{\pi}^k_{, ijk} \,, \label{pisolve} \\
\tilde{\pi}^i &= \Delta^{-1} \pi^{ij}_{~~,j} \,, \\
\hat{\pi}^{ij} &= \frac{1}{2} \left( \delta_{ij} - \partial_i \partial_j \Delta^{-1} \right) \pi^{kk} \,. \label{pihat}
\end{align}
\end{subequations}
This can be shown by inserting (\ref{decompparts}) and (\ref{TTproj}) into
(\ref{decomp}), which then turns into an identity.
We further introduce an alternative vector potential $V^i$ by
\begin{equation}
V^i = \left( \delta_{ij} - \frac{1}{4} \partial_i \partial_j \Delta^{-1} \right) \tilde{\pi}^j \,, \label{Vtrans}
\end{equation}
for which it holds
\begin{equation}
\tilde{\pi}^{ij} = V^i_{, j}
	+ V^j_{, i} - \frac{2}{3} \delta_{ij} V^k_{, k} \,.
\end{equation}
Notice that this decomposition reduces to the one for $\pi^{ij}_{\text{can}}$, as
$\hat{\pi}^{ij}_{\text{can}} = 0$ follows from the gauge condition (\ref{ADMTTpi}).
This gauge condition can also be inserted into (\ref{pihat}) in the form
$\pi^{ii} = - \pi^{ii}_{\text{matter}}$. This immediately yields $\hat{\pi}^{ij}$,
which is zero at the considered order due to $\pi^{ii}_{(5)\text{matter}} = 0$; also see (\ref{tacepia})
for higher orders. Similarly the decomposition of $\gamma_{ij}$ can be derived
using (\ref{ADMTTg}).

Now we have to decide whether to use $\pi^{ij \text{TT}}_{\text{can}}$ and
$\tilde{\pi}^{ij}_{\text{can}}$, or $\pi^{ij \text{TT}}$, $\tilde{\pi}^{ij}$,
and $\hat{\pi}^{ij}$ for the expansion of the field constraints. We choose
the latter option as it simplifies the calculation at the considered PN order.
Then one has to go over to $\pi^{ij \text{TT}}_{\text{can}}$ later on using
\begin{equation}
\pi^{ij\text{TT}} = \pi^{ij\text{TT}}_{\text{can}} - \delta^{\text{TT}ij}_{kl} \pi^{kl}_{\text{matter}} \,.
\end{equation}
Notice that it holds $\tilde\pi^{ij}_{\text{can},j} = \tilde{\pi}^{ij}_{~~,j} + \pi^{ij}_{\text{matter},j}$
and thus $\tilde\pi^{i}_{\text{can}} = \tilde{\pi}^{i} + \partial_j \Delta^{-1} \pi^{ij}_{\text{matter}}$.

The expansion of the momentum constraint immediately follows from the exact formula
\begin{align}
\begin{split}
\tilde{\pi}^{ij}_{~~,j} &=
	- 8\pi \mathcal{H}^{\text{matter}}_{i}
	+ {A^{ij}}_{,j}
	+ B^i
	- \Delta \left( V^k h_{ki}^{\text{TT}} \right) \nl
	+ \frac{1}{2} \pi^{jk\text{TT}} h_{jk,i}^{\text{TT}}
	- ( \pi^{jk\text{TT}} h_{ki}^{\text{TT}} )_{,j} \,,
\end{split}\\
\begin{split}
A^{ij} &= \left[ 1 - \left( 1 + \tfrac{1}{8} \phi \right)^4 \right]
		( \tilde{\pi}^{ij} + \pi^{ij\text{TT}} ) \nl
	+ V^k ( h_{ki,j}^{\text{TT}} + h_{kj,i}^{\text{TT}} - h_{ij,k}^{\text{TT}} )
	- \frac{1}{3} V^k_{,k} h_{ij}^{\text{TT}} \,,
\end{split}\\
B^i &= \frac{1}{2} \hat{\pi}^{jk} \gamma_{jk,i} - \hat{\pi}^{jk} \gamma_{ij,k} \,,
\end{align}
which is analogous to (\ref{cmom}). With the help of
\begin{equation}
\tilde{\pi}^i = \Delta^{-1} \tilde{\pi}^{ij}_{~~,j} \,, \\
\end{equation}
the expanded momentum constraint can be solved iteratively for
$\tilde{\pi}^i$ by applying an inverse Laplacian to it.
$\tilde{\pi}^{ij}$ and $V^i$ then follow from (\ref{pisolve}) and (\ref{Vtrans}).
The expansion of the source $\mathcal{H}^{\text{matter}}_{i}$ is given by (\ref{defP}) and
\begin{align}
	s_{a(3)}^{ij} &= S_{a(i)(j)} \,, \\
	s^{ij}_{a (5)} &= - \frac{1}{2 m_a^2} P_{a k} P_{a i} S_{a(j)(k)}
		+ ( i \leftrightarrow j ) \,, \\
\begin{split}
	s^{ij}_{a (7)} &= \frac{3\vct{P}_a^2}{8m_a^4} P_{a k} P_{a i} S_{a(j)(k)}
		+ \frac{1}{4m_a^2} P_{a k} P_{a i} S_{a(j)(k)} \phi_{(2)} \nl
		- \frac{1}{2} h_{ki}^{\text{TT}} S_{a(j)(k)}
		+ ( i \leftrightarrow j ) \,.
\end{split}
\end{align}
Notice that $s^{ij}_{a (9)}$ is not needed for the Hamiltonian at the considered order,
it only contributes to total linear and angular momentum.

The expansion of the Hamilton constraint (\ref{ham}) reads
\begin{widetext}
\begin{align}
- \frac{1}{16\pi} \Delta \phi_{(2)} &= \mathcal{H}^{\rm matter}_{(2)} \,, \\
- \frac{1}{16\pi} \Delta \phi_{(4)} &= \mathcal{H}^{\rm matter}_{(4)}
	- \frac{1}{8} \mathcal{H}^{\rm matter}_{(2)} \phi_{(2)} \,, \\
- \frac{1}{16\pi} \Delta \phi_{(6)} &= \mathcal{H}^{\rm matter}_{(6)}
	- \frac{1}{8} \left( \mathcal{H}^{\rm matter}_{(4)} \phi_{(2)}
		+ \mathcal{H}^{\rm matter}_{(2)} \phi_{(4)} \right)
	+ \frac{1}{64} \mathcal{H}^{\rm matter}_{(2)} \phi_{(2)}^2
	+ \frac{1}{16\pi} \left[ \left( \tilde{\pi}^{i j}_{(3)} \right)^2
	- \frac{1}{2} \left( \phi_{(2)} h^{\text{TT}}_{i j} \right)_{, i j} \right] \,, \\
\begin{split}
- \frac{1}{16\pi} \Delta \phi_{(8)} &= \mathcal{H}^{\rm matter}_{(8)}
	- \frac{1}{8} \left( \mathcal{H}^{\rm matter}_{(6)} \phi_{(2)}
		+ \mathcal{H}^{\rm matter}_{(4)} \phi_{(4)}
		+ \mathcal{H}^{\rm matter}_{(2)} \phi_{(6)} \right)
	+ \frac{1}{64} \left( \mathcal{H}^{\rm matter}_{(4)} \phi_{(2)}^2
		+ 2 \mathcal{H}^{\rm matter}_{(2)} \phi_{(2)} \phi_{(4)} \right) \nl
	- \frac{1}{512} \mathcal{H}^{\rm matter}_{(2)} \phi_{(2)}^3
	+ \frac{1}{16\pi}\left[ \frac{1}{8} \phi_{(2)} \left(\tilde{\pi}^{i j}_{(3)}\right)^2
	+ 2 \tilde{\pi}^{i j}_{(3)} \tilde{\pi}^{i j}_{(5)}
	- \frac{1}{16} \phi_{(2) , i} \phi_{(2) , j} h^{\text{TT}}_{i j}
	+ \frac{1}{4} \left(h^{\text{TT}}_{i j , k}\right)^2 \right] \nl
	+ \frac{1}{16\pi}\left[ 2 \tilde{\pi}^{i j}_{(3)} \pi^{i j \text{TT}}
	- \frac{1}{2} \left( \phi_{(4)} h^{\text{TT}}_{i j} \right)_{, i j}
	+ \frac{1}{4} \left( \phi_{(2)} \phi_{(2) , j} h^{\text{TT}}_{i j} \right)_{,i}
	- \frac{1}{2} \Delta \left( h^{\text{TT}}_{i j} \right)^2
	+ \frac{1}{2} \left( h^{\text{TT}}_{i j} h^{\text{TT}}_{i k} \right)_{, j k}
	\right] \,,
\end{split}\\
\begin{split}
- \frac{1}{16\pi} \Delta \phi_{(10)} &= \mathcal{H}^{\rm matter}_{(10)}
	- \frac{1}{8} \left( \mathcal{H}^{\rm matter}_{(8)} \phi_{(2)}
		+ \mathcal{H}^{\rm matter}_{(6)} \phi_{(4)}
		+ \mathcal{H}^{\rm matter}_{(4)} \phi_{(6)}
		+ \mathcal{H}^{\rm matter}_{(2)} \phi_{(8)} \right) \nl
	+ \frac{1}{64} \left( \mathcal{H}^{\rm matter}_{(6)} \phi_{(2)}^2
		+ 2 \mathcal{H}^{\rm matter}_{(4)} \phi_{(2)} \phi_{(4)}
		+ 2 \mathcal{H}^{\rm matter}_{(2)} \phi_{(2)} \phi_{(6)}
		+ \mathcal{H}^{\rm matter}_{(2)} \phi_{(4)}^2 \right) \nl
	- \frac{1}{512} \left( \mathcal{H}^{\rm matter}_{(4)} \phi_{(2)}^3
		+ 3 \mathcal{H}^{\rm matter}_{(2)} \phi_{(2)}^2 \phi_{(4)} \right)
	+ \frac{1}{4096} \mathcal{H}^{\rm matter}_{(2)} \phi_{(2)}^4
	- \frac{1}{16} \mathcal{H}^{\rm matter}_{(2)} \left(h^{\text{TT}}_{i j}\right)^2 \nl
	+ \frac{1}{16\pi} \bigg[ \frac{1}{8} \left( \phi_{(4)} \left(\tilde{\pi}^{i j}_{(3)}\right)^2
		+ 2 \phi_{(2)} \tilde{\pi}^{i j}_{(3)} \tilde{\pi}^{i j}_{(5)} \right)
	+ \left( \left(\tilde{\pi}^{i j}_{(5)}\right)^2
		+ 2 \tilde{\pi}^{i j}_{(3)} \tilde{\pi}^{i j}_{(7)} \right)
	+ \frac{1}{4} \phi_{(2)} \tilde{\pi}^{i j}_{(3)} \pi^{i j \text{TT}}
	+ \left(\pi^{i j \text{TT}}\right)^2 \nlq
	+ \left(
		- \frac{1}{8} \phi_{(4) , i} \phi_{(2) , j}
		+ \frac{5}{128} \phi_{(2)} \phi_{(2) , i} \phi_{(2) , j}
		+ 2 \tilde{\pi}^{i k}_{(3)} \tilde{\pi}^{j k}_{(3)}
	\right) h^{\text{TT}}_{i j} \nlq
	- \frac{7}{32} \phi_{(2)} \left(h^{\text{TT}}_{i j , k}\right)^2
	+ \frac{1}{16} \phi_{(2)} \left( h^{\text{TT}}_{i j} h^{\text{TT}}_{i k} \right)_{, j k} \bigg]
	+ (\text{td}) \,,
% 	\nl
% 	+ \frac{1}{16\pi} \bigg[ 2 \tilde{\pi}^{i j}_{(5)} \pi^{i j \text{TT}}
% 		- \frac{1}{2} \left( \phi_{(6)} h^{\text{TT}}_{i j} \right)_{, i j}
% 		+ \frac{1}{4} \left( \phi_{(2)} \phi_{(4) , j} h^{\text{TT}}_{i j} \right)_{,i}
% 		+ \frac{1}{4} \left( \phi_{(4)} \phi_{(2) , j} h^{\text{TT}}_{i j} \right)_{,i} \nlq
% 		- \frac{5}{64} \left( \phi_{(2)}^2 \phi_{(2) , j} h^{\text{TT}}_{i j} \right)_{,i}
% 		+ \frac{7}{16} \Delta \left( \phi_{(2)} \left( h^{\text{TT}}_{i j} \right)^2 \right)
% 		- \frac{1}{2} \left( \phi_{(2)} h^{\text{TT}}_{i j} h^{\text{TT}}_{i k , j} \right)_{,k}
% 		- \frac{1}{4} \left( \phi_{(2),k} \left(h^{\text{TT}}_{i j}\right)^2 \right)_{,k} \bigg]
\end{split}
\end{align}
where $(\text{td})$ denotes a total divergence. These equations can be solved iteratively for $\phi$
by applying an inverse Laplacian to them. The ADM Hamiltonian (\ref{HADM})
results from an integration over the right-hand sides of these equations.
The source expressions are given by
\begin{align}
\mathcal{H}^{\rm matter}_{(2)} &= \sum_a m_a \delta_a \,, \\
\mathcal{H}^{\rm matter}_{(4)} &= \sum_a \left[ \frac{{\bf P}^2_a}{2 m_a} \delta_a
	+ \frac{1}{2 m_a} P_{a i} S_{a (i) (j)} \delta_{a , j} \right] \,, \\
\mathcal{H}^{\rm matter}_{(6)} &= \sum_a \left[
	- \frac{({\bf P}^2_a)^2}{8 m_a^3} \delta_a
	- \frac{{\bf P}^2_a}{4 m_a} \phi_{(2)} \delta_a
	+ \frac{1}{4 m_a} P_{a i} S_{a (i) (j)} \phi_{(2) , j} \delta_a
	- \frac{{\bf P}^2_a}{8 m^3_a} P_{a i} S_{a (i) (j)} \delta_{a , j}
	- \frac{1}{4 m_a} P_{a i} S_{a (i) (j)} ( \phi_{(2)} \delta_a )_{, j} \right] \,, \\
\begin{split}
\mathcal{H}^{\rm matter}_{(8)} &= \sum_a \bigg[
	\frac{({\bf P}^2_a)^3}{16 m^5_a} \delta_a
	+ \frac{({\bf P}^2_a)^2}{8 m^3_a} \phi_{(2)} \delta_a
	+ \frac{5 {\bf P}^2_a}{64 m_a} \phi_{(2)}^2 \delta_a
	- \frac{{\bf P}^2_a}{4 m_a} \phi_{(4)} \delta_a
	- \frac{1}{2 m_a} P_{a i} P_{a j} h^{\text{TT}}_{i j} \delta_a
	- \frac{{\bf P}_a^2}{8 m_a^3} P_{a i} S_{a (i) (j)} \phi_{(2) , j} \delta_a \nlq
	- \frac{5}{32 m_a} P_{a i} S_{a (i) (j)} \phi_{(2)} \phi_{(2) , j} \delta_a
	+ \frac{1}{4 m_a} P_{a i} S_{a (i) (j)} \phi_{(4) , j} \delta_a
	+ \frac{1}{2 m_a} P_{a i} S_{a (j) (k)} h^{\text{TT}}_{i j , k} \delta_a \bigg] \nl
	+ \sum_a \partial_j \bigg[ \frac{({\bf P}^2_a)^2}{16m_a^5} P_{a i} S_{a (i) (j)} \delta_a
	+ \frac{{\bf P}^2_a}{8m_a^3} P_{a i} S_{a (i) (j)} \phi_{(2)} \delta_a
	+ \frac{5}{64m_a} P_{a i} S_{a (i) (j)} \phi_{(2)}^2 \delta_a
	- \frac{1}{4m_a} P_{a i} S_{a (i) (j)} \phi_{(4)} \delta_a \nlq
	+ \frac{1}{4m_a} P_{a i} S_{a(k)(i)} h^{\text{TT}}_{j k} \delta_a
	- \frac{1}{4m_a} P_{a i} S_{a(k)(j)} h^{\text{TT}}_{i k} \delta_a \bigg] \,,
\end{split}\\
\begin{split}
\mathcal{H}^{\rm matter}_{(10)} &= \sum_a \bigg[
	- \frac{5({\bf P}^2_a)^4}{128m_a^7} \delta_a
	- \frac{3({\bf P}^2_a)^3}{32m_a^5} \phi_{(2)} \delta_a
	- \frac{9({\bf P}^2_a)^2}{128m_a^3} \phi_{(2)}^2 \delta_a
	- \frac{5{\bf P}^2_a}{256m_a} \phi_{(2)}^3 \delta_a
	+ \frac{({\bf P}^2_a)^2}{8m_a^3} \phi_{(4)} \delta_a
	+ \frac{5{\bf P}^2_a}{32m_a} \phi_{(2)} \phi_{(4)} \delta_a \nlq
	- \frac{{\bf P}^2_a}{4m_a} \phi_{(6)} \delta_a
	+ \frac{{\bf P}^2_a}{4m_a^3} P_{a i} P_{a j} h^{\text{TT}}_{i j} \delta_a
	+ \frac{1}{2m_a} P_{a i} P_{a j} \phi_{(2)} h^{\text{TT}}_{i j} \delta_a
	+ \frac{3({\bf P}^2_a)^2}{32m_a^5} P_{a i} S_{a (i) (j)} \phi_{(2) , j} \delta_a \nlq
	+ \frac{9{\bf P}^2_a}{64m_a^3} P_{a i} S_{a (i) (j)} \phi_{(2)} \phi_{(2) , j} \delta_a
	+ \frac{15}{256m_a} P_{a i} S_{a (i) (j)} \phi_{(2)}^2 \phi_{(2),j} \delta_a
	- \frac{{\bf P}^2_a}{8m_a^3} P_{a i} S_{a (i) (j)} \phi_{(4) , j} \delta_a \nlq
	- \frac{5}{32m_a} P_{a i} S_{a (i) (j)} \left( \phi_{(2)} \phi_{(4)} \right)_{,j} \delta_a
	+ \frac{1}{4m_a} P_{a i} S_{a (i) (j)} \phi_{(6),j} \delta_a
	- \frac{{\bf P}^2_a}{4m_a^3} P_{a i} S_{a (j) (k)} h^{\text{TT}}_{ij,k} \delta_a \nlq
%	+ \frac{1}{4m_a^3} P_{a i} P_{a j} P_{a k} S_{a (i) (l)} h^{\text{TT}}_{jl,k} \delta_a
	- \frac{1}{2m_a} P_{a i} S_{a (j) (k)} \phi_{(2)}h^{\text{TT}}_{ij,k} \delta_a
	+ \frac{3}{8m_a} P_{a i} S_{a (j) (k)} \phi_{(2),j} h^{\text{TT}}_{ik} \delta_a
	- \frac{1}{8m_a} P_{a i} S_{a (i) (k)} \phi_{(2),j} h^{\text{TT}}_{jk} \delta_a
	\bigg]
	+ (\text{td}) \,.
\end{split}
\end{align}
\end{widetext}

Although the source terms given in this section seem to be SO couplings
only, the expressions given here are enough to give all S$_1$S$_2$ contributions, too.
All S$_1$S$_2$ terms in the Hamiltonian come in from the nonlinearities on the
right-hand sides of the expanded constraints, as in \cite{steinhoff07:spin1,steinhoff08:spin2}.
At higher orders in the single spin variables, however, more contributions are needed
in the source of the constraints (i.e., in the stress-energy tensor).

\section{Field evolution\label{Sevo}}
In this section we derive the wave equation for $h^{\text{TT}}_{i j}$ from
the ADM Hamiltonian and compare it with the corresponding one that follows
directly from the Einstein equations. Agreement is found, which proves that the
constructed ADM Hamiltonian gives the correct time evolution for the gravitational
field up to and including 3.5PN.
This provides a thorough check of the canonical formalism derived in
this paper and also of Ref.\ \cite{SSepl}.
% Further, the formulas derived in this section are useful for various applications.

% The last missing piece to prove that the ADM
% Hamiltonian is correct is to show that it gives the correct time evolution for
% the matter variables, too. However, in order to do so one needs a (3+1)-split
% of the the matter EOM (\ref{MPeqs}), redefine the variables according to
% (\ref{NWvars}) and (\ref{espin}), and perform a PN expansion in the ADMTT gauge.
% Notice that $h^{\text{TT}}_{i j}$ and $\pi^{i j \text{TT}}_{\text{can}}$ do not need
% to be eliminated, as they are still members of the phase space
% of the ADM Hamiltonian, i.e., only the constraints have to be solved and not the
% wave equation for $h^{\text{TT}}_{i j}$. The result should coincide with the matter EOM
% following from the constructed ADM Hamiltonian. This check will not be done here.
% Instead, full consistency is shown by an action approach to the canonical
% formulation at linear order in spin \cite{SSepl}. On the other hand this
% section checks the field evolution of the action approach up to and including 3.5PN.

\subsection{Interaction Hamiltonian and wave equation}
The field EOM can be obtained from the ADM Hamiltonian by
\begin{align}
\frac{1}{16\pi} \dot{h}^{\text{TT}}_{i j} &=
	\delta^{\text{TT} i j}_{k l} \frac{\delta H_{\text{ADM}}}{\delta \pi^{k l \text{TT}}_{\text{can}}} \,, \\
\frac{1}{16\pi} \dot{\pi}^{i j \text{TT}}_{\text{can}} &=
	- \delta^{\text{TT} i j}_{k l} \frac{\delta H_{\text{ADM}}}{\delta h^{\text{TT}}_{k l}} \,,
\end{align}
where the dot denotes a partial time derivative.
It is suitable to introduce an interaction Hamiltonian $H^{\text{int}}$ between matter
and gravitational field as
\begin{equation}
H^{\text{int}} = H_{\text{ADM}}^{\text{TT-parts}} - \frac{1}{16\pi} \int \text{d}^3 x \,
	\left[ \frac{1}{4} (h^{\text{TT}}_{i j , k})^2 + (\pi^{i j \text{TT}}_{\text{can}})^2 \right] \,,
\end{equation}
where $H_{\text{ADM}}^{\text{TT-parts}}$ denotes the parts of the ADM
Hamiltonian depending on $h^{\text{TT}}_{i j}$ and $\pi^{i j \text{TT}}_{\text{can}}$.
The field EOM then read
\begin{subequations}\label{fieldEOM}
\begin{align}
	\frac{1}{16\pi} \Box h^{\text{TT}}_{i j} &= \delta^{\text{TT} i j}_{k l}
			\left[ 2 \frac{\delta H^{\text{int}}}{\delta h^{\text{TT}}_{k l}}
		- \frac{\partial}{\partial t} \frac{\delta H^{\text{int}}}{\delta \pi^{k l \text{TT}}_{\text{can}}} \right] \,, \\
	\frac{1}{16\pi} \pi^{i j \text{TT}}_{\text{can}} &= \frac{1}{2} \left[ \frac{1}{16\pi} \dot{h}^{\text{TT}}_{i j}
		- \delta^{\text{TT} i j}_{k l} \frac{\delta H^{\text{int}}}{\delta \pi^{k l \text{TT}}_{\text{can}}} \right] \,,
\end{align}
\end{subequations}
with $\Box = \Delta - \pa_t^2$ and the partial time derivative $\pa_t$.

Notice that $\phi_{(6)}$, $\phi_{(8)}$, and $\tilde{\pi}^{i j}_{(7)}$ depend
on $h^{\text{TT}}_{ij}$ and/or $\pi^{ij \text{TT}}$, and at a first look
it seems that one has to explicitly solve the constraints for these functions in order
to get the interaction Hamiltonian. However, one can use the expanded constraints to eliminate
$\phi_{(6)}$, $\phi_{(8)}$, and $\tilde{\pi}^{i j}_{(7)}$ by performing
certain partial integrations. With the definitions
\begin{align}
	\phi_{1(4)} &\equiv - 16\pi \Delta^{-1} \mathcal{H}^{\rm matter}_{(4)} \,, \\
	\phi_{2(4)} &\equiv - 16\pi \Delta^{-1}
		\left( - \frac{1}{8} \mathcal{H}^{\rm matter}_{(2)} \phi_{(2)} \right) \,,
\end{align}
for which $\phi_{(4)} = \phi_{1(4)} + \phi_{2(4)}$ holds, these partial integrations read
\begin{align}
	\mathcal{H}^{\rm matter}_{(2)} \phi_{(6)}
		&= \frac{1}{32\pi} \left[ \phi_{(2),i} \phi_{(2),j} h^{\text{TT}}_{i j} \right]
			+ \cdots \,, \\
	\mathcal{H}^{\rm matter}_{(4)} \phi_{(6)}
		&= \frac{1}{32\pi} \left[ \phi_{1(4),i} \phi_{(2),j} h^{\text{TT}}_{i j} \right]
			+ \cdots \,, \label{PIphi6} \\
	\mathcal{H}^{\rm matter}_{(2)} \phi_{(2)} \phi_{(6)}
		&= - \frac{1}{4\pi} \left[ \phi_{2(4),i} \phi_{(2),j} h^{\text{TT}}_{i j} \right]
			+ \cdots \,,
\end{align}
\begin{widetext}
\begin{align}
\begin{split}
	\mathcal{H}^{\rm matter}_{(2)} \phi_{(8)} &= \mathcal{H}^{\rm matter}_{(8)} \phi_{(2)}
		+ \frac{1}{2} \mathcal{H}^{\rm matter}_{(2)} \left(h^{\text{TT}}_{i j}\right)^2
		+ \frac{1}{16\pi} \bigg[ 2 \phi_{(2)} \tilde{\pi}^{i j}_{(3)} \pi^{i j \text{TT}}_{\text{can}}
		+ \frac{1}{2} \phi_{2(4),i} \phi_{(2),j} h^{\text{TT}}_{i j} \nl
		+ \frac{1}{2} \phi_{(4),i} \phi_{(2),j} h^{\text{TT}}_{i j}
		- \frac{5}{16} \phi_{(2)} \phi_{(2) , i} \phi_{(2) , j} h^{\text{TT}}_{i j}
		+ \frac{1}{4} \phi_{(2)} \left(h^{\text{TT}}_{i j , k}\right)^2
		+ \frac{1}{2} \phi_{(2)} \left( h^{\text{TT}}_{i j} h^{\text{TT}}_{i k} \right)_{, j k}
		\bigg] + \cdots \,,
\end{split}\\
\begin{split}
\tilde{\pi}^{i j}_{(3)} \tilde{\pi}^{i j}_{(7)} &= 16\pi V^i_{(3)} \mathcal{H}^{\rm matter}_{(7) i}
	- \frac{1}{2} \phi_{(2)} \tilde{\pi}^{i j}_{(3)} \pi^{i j \text{TT}}_{\text{can}}
	+ \left( - 2 \tilde{\pi}^{i k}_{(3)} V^j_{(3),k}
		+ \tilde{\pi}^{i j}_{(3),k} V^k_{(3)}
	+ \frac{3}{4} \tilde{\pi}^{i j}_{(3)} \tilde{\pi}^k_{(3),k}
	\right) h_{ij}^{\text{TT}}
	+ \cdots \,,
\end{split}
\end{align}
where dots denote total divergences and/or terms independent
of $h^{\text{TT}}_{ij}$ and $\pi^{ij \text{TT}}_{\text{can}}$. The interaction
Hamiltonian then results as
\begin{equation}
\begin{split}
H^{\text{int}} &= \frac{1}{16\pi} \int \text{d}^3 x \, \bigg[
	\left( B_{(4)ij} + \hat{B}_{(6)ij} \right) h^{\text{TT}}_{i j}
	- \frac{16\pi}{8} \mathcal{H}^{\rm matter}_{(2)} \left(h^{\text{TT}}_{i j}\right)^2
	- \frac{1}{4} \phi_{(2)} \left(h^{\text{TT}}_{i j , k}\right)^2
	+ 2(  V^i_{(3)} \phi_{(2),j} - \pi^{i j}_{(5)\text{matter}} ) \pi^{i j \text{TT}}_{\text{can}}
	\bigg] \,,
\end{split}
\end{equation}
where
\begin{align}
B_{(4)ij} &= 16\pi \frac{\delta \left(
		\int{ d^3 x \, \mathcal{H}^{\rm matter}_{(8)}  } \right)}
		{\delta h^{\text{TT}}_{i j}}
	- \frac{1}{8} \phi_{(2) , i} \phi_{(2) , j} \,, \\
\begin{split}
\hat{B}_{(6)ij} &= 16\pi \frac{\delta \left(
		\int{ d^3 x \, \left( \mathcal{H}^{\rm matter}_{(10)}
			- \frac{1}{4} \mathcal{H}^{\rm matter}_{(8)} \phi_{(2)}
			+ 2 \mathcal{H}^{\rm matter}_{(7) k} V^k_{(3)} \right) } \right)}
		{\delta h^{\text{TT}}_{ij}}
	+ \frac{1}{4} \phi_{1(4)} \phi_{(2) , i j}
	+ \frac{3}{8} \phi_{2(4)} \phi_{(2) , i j} \nl
	+ \frac{5}{64} \phi_{(2)} \phi_{(2) , i} \phi_{(2) , j}
	+ 2 \tilde{\pi}^{j k}_{(3)} \left( \tilde{\pi}^k_{(3),i} - \tilde{\pi}^i_{(3),k} \right)
	+ 2 \tilde{\pi}^{i j}_{(3),k} V^k_{(3)}
	+ \frac{1}{2} \tilde{\pi}^{i j}_{(3)} \tilde{\pi}^k_{(3),k} \,,
\end{split}
\end{align}
The field EOM finally result from (\ref{fieldEOM}) as
\begin{align}
\Box h^{\text{TT}}_{i j} &= \delta^{\text{TT} i j}_{k l}
	\left[ 2 B_{(4)kl} + 2 B_{(6)kl}
	- \frac{16\pi}{2} \mathcal{H}^{\rm matter}_{(2)} h^{\text{TT}}_{kl}
	+ \left( \phi_{(2)} h^{\text{TT}}_{kl,m} \right)_{,m}
	- 2 \frac{d}{d t} \left( V^k_{(3)} \phi_{(2),l} \right) \right] \,, \label{wave} \\
\pi^{i j \text{TT}}_{\text{can}} &= \frac{1}{2} \dot{h}^{\text{TT}}_{i j}
	- \delta^{\text{TT} i j}_{k l}  \left( V^k_{(3)} \phi_{(2),l} - \pi^{kl}_{(5)\text{matter}} \right) \,, \label{httdot}
\end{align}
with $B_{(6)ij} = \hat{B}_{(6)ij} + \dot{\pi}^{ij}_{(5)\text{matter}}$.
For our source, one gets
\begin{align}
B_{(4)ij} &= 16\pi \sum_a \left[ - \frac{1}{2 m_a} P_{a i} P_{a j} \delta_a
	- \frac{1}{2 m_a} P_{a i} S_{a (j) (k)} \delta_{a,k} \right]
	- \frac{1}{8} \phi_{(2) , i} \phi_{(2) , j} \,, \\
\begin{split}
B_{(6)ij} &= 16\pi \sum_a
	\bigg[ \frac{{\bf P}^2_a}{4m_a^3} P_{a i} P_{a j} \delta_a
	+ \frac{5}{8m_a} P_{a i} P_{a j} \phi_{(2)} \delta_a
	+ \frac{{\bf P}^2_a}{4m_a^3} P_{a i} S_{a (j) (k)} \delta_{a,k}
	- \frac{1}{4m_a^3} P_{a l} P_{a j} P_{a k} S_{a (l) (i)} \delta_{a,k} \nlq
	+ \frac{5}{8m_a} P_{a i} S_{a (j) (k)} \left( \phi_{(2)} \delta_a \right)_{,k}
	+ \frac{1}{2m_a} P_{a i} S_{a (k) (j)} \phi_{(2),k} \delta_a
	- \frac{1}{8m_a} P_{a k} S_{a (k) (i)} \phi_{(2),j} \delta_a \nlq
	+ \frac{1}{2} S_{a (k) (i)} \left( V^j_{(3),k} + V^k_{(3),j} \right) \delta_a \bigg] \nl
	+ \frac{1}{2} \phi_{1(4)} \phi_{(2) , i j}
	+ \frac{3}{8} \phi_{2(4)} \phi_{(2) , i j}
	+ \frac{5}{64} \phi_{(2)} \phi_{(2) , i} \phi_{(2) , j}
	+ 2 \tilde{\pi}^{j k}_{(3)} \left( \tilde{\pi}^k_{(3),i} - \tilde{\pi}^i_{(3),k} \right)
	+ 2 \tilde{\pi}^{i j}_{(3),k} V^k_{(3)}
	+ \frac{1}{2} \tilde{\pi}^{i j}_{(3)} \tilde{\pi}^k_{(3),k} \,.
\end{split}
\end{align}
\end{widetext}
Here the $\phi_{(6)}$ terms in $\mathcal{H}^{\rm matter}_{(10)}$
were rewritten as $-\frac{1}{2} \mathcal{H}^{\rm matter}_{(4)} \phi_{(6)} + (\text{td})$
and handled by (\ref{PIphi6}). The time derivative $\dot{\pi}^{ij}_{(5)\text{matter}}$
was calculated using leading-order Hamiltonians.
Notice that the formula given for the interaction Hamiltonian (and thus the wave equation)
is quite general and in principle applicable not only to linear order in spin.
Further, for nonspinning objects the result in \cite{JS97} is reproduced.

One can remove $\pi^{i j}_{(5)\text{matter}}$ from the
interaction Hamiltonian by a canonical transformation generated by
\begin{equation}
	g = \frac{1}{16\pi} \int \text{d}^3 x \, \pi^{i j}_{(5)\text{matter}} h^{\text{TT}}_{i j} \,,
\end{equation}
corresponding to the choice $C=1$ in Eqs.\ (\ref{Cambig}).

\subsection{Comparison with the Einstein equations}
The time evolution parts of the Einstein equations in the variables
used here read
\begin{align}
	\dot{\gamma}_{ij} &= 2 N \gamma^{-1/2}
		(\pi_{ij} - \tfrac{1}{2} \gamma_{ij} \gamma_{kl} \pi^{kl})
		+ N_{i ; j} + N_{j ; i}\,, \\
\begin{split}
	\dot{\pi}^{ij} &= - N \sqrt{\gamma} (\text{R}^{ij} - \tfrac{1}{2} \gamma^{ij} \text{R}) \nl
		+ \tfrac{1}{2} N \gamma^{-1/2} \gamma^{ij} (\pi^{mn} \pi_{mn}
		- \tfrac{1}{2} (\gamma_{mn} \pi^{mn})^2) \nl
		- 2 N \gamma^{-1/2} (\gamma_{mn} \pi^{im} \pi^{nj}
		- \tfrac{1}{2} \gamma_{mn} \pi^{mn} \pi^{ij}) \nl
		+ \sqrt{\gamma} (N^{; ij} - \gamma^{ij} N^{;m}_{~~~ ;m} )
		+ (\pi^{ij} N^m)_{;m} \nl
		- N^i_{~;m} \pi^{mj} - N^j_{~ ;m} \pi^{mi}
		+ 8\pi N \gamma^{im} \gamma^{jn} \mathcal{T}_{mn} \,.
\end{split}
\end{align}
(Notice that there are misprints in Eq.\ (8.4) in \cite{steinhoff08:spin2}.)
After using constraints and coordinate conditions, this can be compared to
the results of the last section. Lapse and shift are fixed by the requirement
that the gauge conditions (\ref{ADMTT}) are preserved in time.
From $\delta_{ij} \dot{\pi}^{ij} = 0$ it follows
\begin{align}
	\Delta N_{(2)} &= \frac{1}{4} \mathcal{H}^{\text{matter}}_{(2)} \,, \\
\begin{split}
	\Delta N_{(4)} &= 4\pi \mathcal{T}_{(4) ii} + 4\pi \mathcal{H}^{\text{matter}}_{(4)} \nl
			- \pi \mathcal{H}^{\text{matter}}_{(2)} \phi_{(2)}
			+ \frac{1}{32} \Delta \phi_{(2)}^2 \,,
\end{split}
\end{align}
with solutions
\begin{align}
	N_{(0)} &= 1 \,, \qquad
	N_{(2)} = - \frac{1}{4} \phi_{(2)} \,, \\
	N_{(4)} &= 4\pi \Delta^{-1} \mathcal{T}_{(4) ii} - \frac{1}{4} \phi_{1(4)}
		- \frac{1}{2} \phi_{2(4)} + \frac{1}{32} \phi_{(2)}^2 \,,
\end{align}
while $\dot{\gamma}_{ij,j} - \tfrac{1}{3} \dot{\gamma}_{jj,i} = 0$ leads to
\begin{equation}
	\Delta N_{(3) i} + \frac{1}{3} N_{(3) j , j i} =  16\pi \mathcal{H}^{\text{matter}}_{(3) i} \,, \\
\end{equation}
with the solution
\begin{equation}
	N_{(3) i} = - 2 V_{(3)}^i \,.
\end{equation}

Using the Hamilton constraint and these expressions for lapse and shift, the
PN expansion of the TT-projected evolution equations reads
\begin{align}
	\dot{h}^{\text{TT}}_{i j} &= 2 \pi^{\text{TT} i j} + 2 \delta^{\text{TT} i j}_{k l} ( V_{(3)}^k \phi_{(2) ,l} )
		\,, \label{httdot2} \\
\begin{split}
\dot{\pi}^{i j \text{TT}} &= \frac{1}{2} \Delta h^{\text{TT}}_{i j}
	- \delta^{\text{TT} i j}_{k l} \bigg[ B_{(4)kl} + B_{(6)kl} \nl
	- 4\pi \mathcal{H}^{\rm matter}_{(2)} h^{\text{TT}}_{kl}
	+ \frac{1}{2} \left( \phi_{(2)} h^{\text{TT}}_{kl,m} \right)_{,m} \bigg] \,,
\end{split}
\end{align}
with
\begin{align}
B_{(4)ij} &= - 8\pi \mathcal{T}_{(4) ij}
	- \frac{1}{8} \phi_{(2) , i} \phi_{(2) , j} \,, \\
\begin{split}
B_{(6)ij} &= - 8\pi \mathcal{T}_{(6) ij}
	+ 10\pi \mathcal{T}_{(4) ij} \phi_{(2)}
	- 2 \pi \phi_{(2) , i j} \Delta^{-1} \mathcal{T}_{(4) kk} \nl
	+ \frac{1}{4} \phi_{1(4)} \phi_{(2) , i j}
	+ \frac{3}{8} \phi_{2(4)} \phi_{(2) , i j} \nl
	+ \frac{5}{64} \phi_{(2)} \phi_{(2) , i} \phi_{(2) , j}
	+ 2 \tilde{\pi}^{j k}_{(3)} \left( \tilde{\pi}^k_{(3),i} - \tilde{\pi}^i_{(3),k} \right) \nl
	+ 2 \tilde{\pi}^{i j}_{(3),k} V^k_{(3)}
	+ \frac{1}{2} \tilde{\pi}^{i j}_{(3)} \tilde{\pi}^k_{(3),k} \,.
\end{split}
\end{align}
This can be written as a wave equation identical to (\ref{wave}),
but now with different expressions for $B_{(4)ij}$ and $B_{(6)ij}$.
Also, Eq.\ (\ref{httdot2}) obviously is the same as (\ref{httdot}).
The question thus is if the results for $B_{(4)ij}$ and $B_{(6)ij}$
coincide with the ones from the last section. To see this we need
the expressions for $\mathcal{T}_{(4) ij}$ and $\mathcal{T}_{(6) ij}$.
These follow from (\ref{Tij}) after the transition to NW variables by
(\ref{NWvars}) and PN expansion as
\begin{widetext}
\begin{align}
\mathcal{T}_{(4) i j} &= \sum_a \frac{1}{m_a} P_{a i} P_{a j} \delta_a
	+ \sum_a \frac{1}{2 m_a} \partial_k \Bigg[ P_{a i} S_{a (j)(k)} \delta_a
		+ P_{a j} S_{a (i)(k)} \delta_a \Bigg] \,, \\
\begin{split}
\mathcal{T}_{(6) i j} &= \sum_a \Bigg[ - \frac{\vct{P}_a^2}{2 m_a^3} P_{a i} P_{a j} \delta_a
	- \frac{1}{2} S_{a (k)(i)} \tilde{\pi}^{kj}_{(3)} \delta_a
	- \frac{1}{2} S_{a (k)(j)} \tilde{\pi}^{ki}_{(3)} \delta_a
	+ \frac{1}{8m_a} P_{a i} S_{a (k)(j)} \phi_{(2),k} \delta_a \nlq
	+ \frac{1}{8m_a} P_{a j} S_{a (k)(i)} \phi_{(2),k} \delta_a
	+ \frac{1}{8m_a} P_{a k} S_{a (k)(i)} \phi_{(2),j} \delta_a
	+ \frac{1}{8m_a} P_{a k} S_{a (k)(j)} \phi_{(2),i} \delta_a \Bigg] \nl
	+ \sum_a \frac{1}{4 m_a^3} \partial_k \Bigg[
		\vct{P}_a^2 P_{a i} S_{a (k)(j)} \delta_a
		+ \vct{P}_a^2 P_{a j} S_{a (k)(i)} \delta_a
		+ P_{a i} P_{a k} P_{a l} S_{a (l)(j)} \delta_a
		+ P_{a j} P_{a k} P_{a l} S_{a (l)(i)} \delta_a \Bigg] \,,
\end{split}
\end{align}
\end{widetext}
For our source it holds $\mathcal{T}_{(4) ii} = 2 \mathcal{H}^{\rm matter}_{(4)}$ and
thus $\Delta^{-1} \mathcal{T}_{(4) ii} = - \frac{1}{8 \pi} \phi_{1(4)}$.
The quantities $B_{(4)ij}$ and $B_{(6)ij}$, and thus the evolution
equations of the gravitational field, can now be seen to coincide
with the result of the last section.

\section{Energy flux\label{Sflux}}
In this section we reproduce, within the ADMTT gauge, the 1PN
energy flux at the SO level, obtained in \cite{kidder92,kidder95}
within the harmonic gauge. This should be seen as a further check for the wave equation
(with source terms depending on standard canonical variables) derived
in the last section, which is most important for the calculation
of the 3.5PN Hamiltonian. It also gives a check of the applied regularization techniques.
% Whereas the last section confirmed
% the canonicity of the field variables, this section gives a
% check that the matter variables are canonical (though at a rather low order),
% as the time derivatives appearing in this section are obtained
% in a canonical way (as Poisson brackets with the Hamiltonian).
Notice that the Newtonian flux in the
SO and S$_1$S$_2$ cases vanishes. We will restrict to two
objects here.

\subsection{Far zone expansion of the wave equation}
The retarded solution of the wave equation reads
\begin{equation}
	\Box^{-1}_{\text{ret}} f (\vct{x}, t) \equiv - \frac{1}{4\pi} \int d^3 x^{\prime}
		\frac{f( \vct{x}^{\prime}, t_{\text{ret}} )}{| \vct{x} - \vct{x}^{\prime} |} \,,
\end{equation}
where $f$ is some field and $t_{\text{ret}} = t - c^{-1} | \vct{x} - \vct{x}^{\prime} |$.
We explicitly show the speed of light $c$ here in order to simplify the discussion.
$f$ shall not change much during time intervals significantly smaller than some time interval $T$,
e.g., $f$ describes a binary system with an orbital period $T$ and the
internal dynamics of the individual objects does not introduce
a significantly smaller time scale relevant for $f$.
This allows Taylor expansion in time of $f$ in certain cases.
The first case is the near zone defined by $| \vct{x} - \vct{x}^{\prime} | \ll cT$ where
$f( \vct{x}^{\prime}, t_{\text{ret}} )$, and thus $\Box^{-1}_{\text{ret}} f$,
can formally be expanded in $c^{-1}$. The near zone expansion
is important for the calculation of the Hamiltonian, as the metric
at the position of the spinning objects is needed there.
The second case is the far zone (or wave zone) defined by $| \vct{x}^{\prime} | \ll cT \ll |\vct{x}| \equiv R$
(if the support of $f$ is centered around the origin).
Again we can formally expand in $c^{-1}$, but the quantity
$t_{\text{ret}}^{\text{fz}} = t - \frac{R}{c}$ must be held constant.
Both near and far zone expansion thus fit well into the PN scheme,
as they can be seen as expansions in $c^{-1}$.
Useful formulas for the far zone expansion are
\begin{align}
	| \vct{x} - \vct{x}^{\prime} | &= R - \vct{n} \cdot \vct{x}^{\prime} + \Order{( R^{-1} )} \,, \\
	\frac{1}{| \vct{x} - \vct{x}^{\prime} |} &= \frac{1}{R} + \Order{( R^{-2} )} \,, \\
% \partial_L \frac{f( \vct{x}^{\prime}, t_{\text{ret}}^{\text{fz}} )}{R} &=
% 	\frac{(-)^l n^L}{R} \partial_{ct}^l f( \vct{x}^{\prime}, t_{\text{ret}}^{\text{fz}} ) + \Order{( R^{-2} )} \,, \label{fzder} \\
\partial_L \Delta^{-\frac{l}{2}} \frac{f( \vct{x}^{\prime}, t_{\text{ret}}^{\text{fz}} )}{R}
	&= \frac{n^L}{R} f( \vct{x}^{\prime}, t_{\text{ret}}^{\text{fz}} ) + \Order{( R^{-2} )} \,. \label{fzTT}
\end{align}
where $n^i = x^i / R$ and $L$ is a multi-index with $l$ even.
Notice that the spacial derivative of $t_{\text{ret}}^{\text{fz}}$ does not vanish.
Equation (\ref{fzTT}) can be shown using
\begin{align}
\Delta^{-\frac{l}{2}} R^{-1} e^{-\frac{i\omega}{c} R} &= \frac{1}{(\tfrac{i\omega}{c})^l R} \Bigg[ e^{-\frac{i\omega}{c} R}
	- \sum_{j=0}^{l-1} \frac{(\tfrac{-i\omega}{c} R)^j}{j!} \Bigg] \,,
\end{align}
and the Fourier transform of $f( \vct{x}^{\prime}, t_{\text{ret}}^{\text{fz}} )$
with respect to time,
\begin{equation}
f( \vct{x}^{\prime}, t_{\text{ret}}^{\text{fz}} ) = \int d^3 \omega
	f( \vct{x}^{\prime}, \omega ) e^{i \omega t_{\text{ret}}^{\text{fz}}} \,.
\end{equation}
This Fourier transform can also be used to show
\begin{equation}
\Box^{-1}_{\text{ret}} f = - \frac{1}{4\pi R} \int d^3 x^{\prime} e^{c^{-1} \vct{n} \cdot \vct{x}^{\prime} \partial_{t}}
	f( \vct{x}^{\prime}, t_{\text{ret}}^{\text{fz}} ) + \Order{( R^{-2} )} \,,
\end{equation}
from which the far zone expansion in $c^{-1}$ can most easily be obtained; one just needs to
plug in the Taylor series of $e^{c^{-1} \vct{n} \cdot \vct{x}^{\prime} \partial_{t}}$ to
the required order. This is precisely the multipole expansion of the far zone field.

Now we write the wave equation for $h^{\text{TT}}_{i j}$, Eq.\ (\ref{wave}), in the form
\begin{equation}
\Box h^{\text{TT}}_{i j} = - 8\pi \delta^{\text{TT} i j}_{k l} S_{kl} \,.
\end{equation}
We can replace $S_{kl}$ by its STF part
\begin{equation}
S_{kl}^{\text{STF}} = \frac{1}{2} (S_{kl} + S_{lk}) - \frac{1}{3} \delta_{kl} S_{ii} \,,
\end{equation}
here. The 1PN far zone expansion of
\begin{equation}
h^{\text{TT}}_{i j} = - 8\pi \delta^{\text{TT} i j}_{k l} \Box^{-1}_{\text{ret}} S_{kl}^{\text{STF}} \,,
\end{equation}
results in (from now on $c$ is dropped again)
\begin{equation}
\begin{split}
h^{\text{TT}}_{i j} &= \frac{2}{R} P_{ijkl} \bigg[ I_{kl}(t_{\text{ret}}^{\text{fz}})
		+ n^m \dot{I}_{klm}(t_{\text{ret}}^{\text{fz}}) \nl
		+ \frac{n^m n^n}{2} \ddot{I}_{klmn}(t_{\text{ret}}^{\text{fz}})
	\bigg] + \Order{(R^{-2})} \label{fzhtt} \,,
\end{split}
\end{equation}
with the multipole moments
\begin{equation}
I_{klM}(t) = \int d^3 x^{\prime} x^{\prime M} S_{kl}^{\text{STF}}(\vct{x}^{\prime}, t) \,.
\end{equation}
Here $M$ is a multi-index and the moments $I_{klM}$ are STF with respect to $k$ and $l$.
At higher orders it is better to work with multipole moments which are STF in all indices; see, e.g.,
\cite{blanchet:2pnso2,Damour:Iyer:1991} and references therein.
The TT-projector $\delta^{\text{TT} i j}_{k l}$ was replaced by $P_{ijkl}$,
\begin{align}
P_{ijkl} &= P_{i(k} P_{l)j} - \frac{1}{2} P_{ij} P_{kl} \,, \\
P_{ij} &= \delta_{ij} - n^i n^j \,,
\end{align}
using (\ref{fzTT}). The integrations needed for the multipole
moments also appear in the calculation of the Hamiltonian.
Details on the calculation and the applied regularization techniques
will be given in \cite{part2}, we only show the results here.
Though the expressions for the multipole moments are quite
long, after extracting total time derivatives as
\begin{align}
I_{ij} &= \ddot{Q}_{ij} \,, \\
I_{ijk} &= \dot{Q}_{ijk} \,, \\
I_{ijkl} &= Q_{ijkl} \,,
\end{align}
they can be written in the compact form
\begin{align}
\begin{split}
Q_{ij} &= \bigg[ m_1 \hat{z}_1^i \hat{z}_1^j - \frac{1}{m_1} P_{1k} S_{1(k)(i)} \hat{z}_1^j
	\bigg]_{\text{STF}_{ij}} \nl + (1 \leftrightarrow 2) \,,
\end{split}\\
\begin{split}
Q_{ijk} &= \bigg[ 2 P_{1i} \hat{z}_1^j \hat{z}_1^k - \hat{z}_1^i \hat{z}_1^j P_{1k} - 2 \hat{z}_1^i S_{1(j)(k)}
	\bigg]_{\text{STF}_{ij}} \nl + (1 \leftrightarrow 2) \,,
\end{split}\\
\begin{split}
Q_{ijkl} &= \bigg[ - \frac{2}{m_1} P_{1i} ( S_{1(j)(k)} \hat{z}_1^l + S_{1(j)(l)} \hat{z}_1^k )
	\bigg]_{\text{STF}_{ij}} \nl + (1 \leftrightarrow 2) \,.
\end{split}
\end{align}
Here the subscript $\text{STF}_{ij}$ means to take the STF part in $i$ and $j$, and
$(1 \leftrightarrow 2)$ denotes an exchange of the particle labels.
Only terms needed for the 1PN SO part of the flux are shown.
As the source of the wave equation is expressed in terms of variables with a standard
canonical meaning, leading order Hamiltonians are used to calculate the
time derivatives appearing here.

\subsection{1PN energy flux}
The energy flux $\mathcal{L}$ results from the formula
\begin{equation}
\mathcal{L} = \frac{1}{32\pi} \lim_{R \rightarrow \infty} R^2 \oint d \Omega
	\dot{h}^{\text{TT}}_{i j} \dot{h}^{\text{TT}}_{i j} \,,
\end{equation}
where $t_{\text{ret}}^{\text{fz}}$ is held constant in the limit $R \rightarrow \infty$.
Using (\ref{fzhtt}) we can express this in terms of $I_{kl}$, $I_{klm}$, and $I_{klmn}$ as
\begin{equation}
\begin{split}
\mathcal{L}_{\text{1PN}} &= \frac{1}{5} (\dot{I}_{ij})^2 + \frac{1}{35 c^2} \bigg(
	\frac{11}{3} (\ddot{I}_{ijk})^2 - 2 \ddot{I}_{ijk} \ddot{I}_{ikj} \nl
	- 2 (\ddot{I}_{ikk})^2 - 4 \dot{I}_{ij} {\stackrel{(3)}{I}\!}_{ikjk} + \frac{11}{3} \dot{I}_{ij} {\stackrel{(3)}{I}\!}_{ijkk} \bigg) \,.
\end{split}
\end{equation}
A symbol $(n)$ on top of a multipole denotes the $n$-th time derivative.
Plugging in our expressions for the multipole moments we get for the SO part
\begin{equation}
\begin{split}
\mathcal{L}_{\text{SO}} &=
	\frac{8 M^2 \nu}{15 r^6} \vct{L} \cdot \vct{S}_1 \bigg[
		\left( 27 \dot{r}^2 - 37 \vct{v}^2 - \frac{12 M}{r} \right) \nl
		+ \rho_{21} \left( 18 \dot{r}^2 - 19 \vct{v}^2 - \frac{8 M}{r} \right)
	\bigg] + (1 \leftrightarrow 2) \,,
\end{split}
\end{equation}
where $\vct{S}_a$ has components $S_{a (i)}$,
% and for the S$_1$S$_2$ part
% \begin{equation}
% \begin{split}
% \mathcal{L}_{\text{S$_1$S$_2$}} &= \dots \,,
% \end{split}
% \end{equation}
after going to the center-of-mass frame (see Appendix \ref{COM}), where it holds
\begin{align}
M &= m_1 + m_2 \,, \\
\rho_{21} &= \frac{m_2}{m_1} = \rho_{12}^{-1} \,, \\
\nu &= \frac{\rho_{21}}{(1+\rho_{21})^2} = \frac{\rho_{12}}{(1+\rho_{12})^2} \,, \\
\vct{r}_{12} &= \hat{\vct{z}}_1 - \hat{\vct{z}}_2 \,, \quad
\vct{v} = \dot{\hat{\vct{z}}}_1 - \dot{\hat{\vct{z}}}_2 \\
r &= \| \vct{r}_{12} \| \,, \quad \vct{n}_{12} = \frac{\vct{r}_{12}}{r} \,, \\
\vct{p} &= \vct{P}_1 = - \vct{P}_2 \,, \\
\vct{L} &= \vct{r}_{12} \times \vct{p} \,.
\end{align}
Our result for $\mathcal{L}_{\text{SO}}$
exactly coincides with the one in \cite{kidder92,kidder95}.
% This energy flux will be compared with the instantaneous
% energy loss in the ADM gauge in \cite{part2}.
% The leading order point-mass (PM) contribution in these variables simply is
% \begin{equation}
% \mathcal{L}_{\text{PM}} = \frac{8 M^4 \nu^2}{15 r^4} ( 12 \vct{v}^2 - 11 \dot{r}^2 ) \,.
% \end{equation}

Similarly, we could get the total angular momentum flux.
This requires one to keep all $\Order{(R^{-2})}$ terms.

\section{Conclusions and outlook\label{Sout}}
In the present paper we extend the ADM canonical formalism to spinning objects
up to and including 3.5PN and linear in the single spin variables. Further, general
formulas for the interaction Hamiltonian and the wave equation for $h^{\text{TT}}_{i j}$ are derived.
This is the foundation for the calculation of the 3.5PN SO and S$_1$S$_2$
radiation-reaction Hamiltonians in \cite{part2}
(which, respectively, are 4PN and 4.5PN for maximally rotating black holes),
as they result from the general interaction Hamiltonian by utilizing a
near zone expansion of the wave equation \cite{JS97}.
The important difference to the formalism in \cite{steinhoff08:spin2}
is the spin-dependent correction to the canonical field momentum.

The conservative NNLO SO and S$_1$S$_2$ Hamiltonians
(which are formally at 3PN or, respectively, at 3.5PN and 4PN for maximally rotating black holes)
can also be obtained from the results of the present paper, once they have
been rederived in arbitrary dimension. This is needed for dimensional
regularization, which is the only one known to work consistently
at 3PN in the ADM formalism; see \cite{DJS01,DJS08b}.
The NNLO SO Hamiltonian is the last missing piece to complete the
EOM for maximally rotating binary black holes up to and including 3.5PN.

As the calculation of the mentioned NNLO Hamiltonians is quite involved
(comparable to the 3PN point-mass Hamiltonian), it is a good idea to
thoroughly check the used formalism before starting such a calculation.
The present paper provides several such checks. The wave equation for $h^{\text{TT}}_{i j}$
was compared with the Einstein equations and applied to the leading
order SO energy flux.
% The latter also provides a check of the applied regularization techniques.
Further, agreement with the action approach in \cite{SSepl}
up to the considered order was shown.

The method to construct a canonical formalism in the present
paper is formulated in a quite general way and could be applied also in other situations.
In particular, our method makes no use of a covariant
generalization of flat-space expressions, in contrast to \cite{steinhoff08:spin2}.
It is thus applicable to nonminimal couplings, which appear at the S$_1^2$ level.
One can use the four-dimensional stress-energy tensor with S$_1^2$ quadrupole terms
\cite{SHS08b,Steinhoff:Puetzfeld:2009} to derive the NLO S$_1^2$ Hamiltonian
with the method of the present paper. This Hamiltonian was already
obtained in \cite{SHS08b,HS08}, and agreement with the spin EOM from \cite{PR08b}
was shown in \cite{SS09}. The LO conservative dynamics was obtained in \cite{barker75:so,barker79:ss}.
For more LO results at the S$_1^2$ level oriented toward application in GW astronomy, see, e.g.,
\cite{Poisson:1998,Gergely:Keresztes:2003,Keresztes:Mikoczi:Gergely:2005,Flanagan:Hinderer:2007,Racine:2008}.

\acknowledgments
We gratefully acknowledge many useful discussions with G.\ Sch\"afer.
We further thank P.\ Jaranowski for sharing his insight in the
calculation of the 3.5PN point-mass Hamiltonian.
JS wishes to thank S.\ Hergt and M.\ Tessmer for useful discussions.
HW thanks J.\ Zeng for helpful discussions.
This work is supported by the Deutsche Forschungsgemeinschaft (DFG) through
SFB/TR7 ``Gravitational Wave Astronomy'' and GRK 1523.

\appendix

\section{PN orders and spin\label{Scount}}
In the present paper, PN orders (orders in $c^{-2}$) are counted in terms of the
velocity of light $c$ originally present in the Einstein equations.
We call this formal counting. This has some computational advantages in our method,
e.g., similarities to calculations for nonspinning objects are more manifest. Further, this
formal way of counting best reflects the computational demands, e.g., the
difficulty of the integrations and the regularization techniques that need to be applied.

On the other hand, it makes sense to assume that the spin variables possess
a numerical value of the order $c^{-1}$, which holds for maximally rotating black holes.
This way of counting best reflects the relevance of the spin corrections
to the motion of rapidly rotating objects. Compared to the formal counting, it adds half a PN
order for each spin variable appearing in a specific expression. For example,
the NLO SO and S$_1$S$_2$ Hamiltonians are of the orders 2.5PN and 3PN for
rapidly rotating objects, respectively, but are both of the order 2PN in the formal counting.

To conclude, the formal counting overestimates the importance of the
spin corrections, while the other way of counting overestimates
the computational complexity.

\section{More on the Poincar\'e algebra\label{poincare}}
The action of an element of the three-dimensional Euclidean group, a subgroup of the Poincar\'e group,
on the coordinates of the three-dimensional hypersurfaces can be written as
\begin{equation}
	x^i \rightarrow R_{ij}(\omega) ( x^j + a^j ) \,, \\
\end{equation}
with $a^i$ a constant infinitesimal vector describing a translation
and a rotation matrix $R_{ij}(\omega)$.
It holds $R(\omega) = e^{\omega}$, where $\omega^{ij} = \omega^{ji}$ is a constant
antisymmetric matrix describing the axis and angle of the rotation.
This is the standard representation of the Euclidean group on the coordinates.
On a field, e.g., the metric, the standard representation of the Euclidean groups acts as
\begin{equation}\label{fieldt}
\gamma_{ij}(\vct{x}) \rightarrow R_{ik}(\omega) R_{jl}(\omega) \gamma_{kl}(R^{-1}(\omega) \vct{x} - \vct{a}) \,.
\end{equation}
Obviously, the ADMTT gauge conditions (\ref{ADMTT}) manifestly respect the
Euclidean group in its standard representation. Thus the global Euclidean group,
as a part of the global Poincar\'e group,
is given by its standard representation in the ADMTT gauge.

Now we restrict to infinitesimal transformations, i.e.,
$a^i$ and $\omega^{ij}$ shall be small. Then it holds
\begin{subequations}\label{eucl}
\begin{align}
	x^k &\rightarrow x^k + \delta x^k \,, \\
	\delta x^k &= \tfrac{1}{2} \omega^{ij} M_{ij}^{kl} x^l + a^k \,,
\end{align}
\end{subequations}
where $M_{ij}^{kl} = \delta_i^k \delta_j^l - \delta_j^k \delta_i^l$. The matrices
$M_{ij}$ with components $(M_{ij})^{kl} \equiv M_{ij}^{kl}$ satisfy the commutation relations
\begin{align}
[ M_{ij} , M_{kl} ] &= \delta_{ik} M_{jl} - \delta_{jk} M_{il} + \delta_{il} M_{kj} - \delta_{jl} M_{ki} \\
	&= M_{ij}^{km} M_{ml} + M_{ij}^{lm} M_{km} \,.
\end{align}
Thus the matrices $M_{ij}$ form a representation of the Lie-algebra $\text{so}(3)$,
namely, the vector representation. Further from (\ref{fieldt}) we have
\begin{equation}
\begin{split}
\gamma_{ij} &\rightarrow \gamma_{ij} - \tfrac{1}{2} \omega^{kl} M_{kl}^{mn} x^n \partial_m \gamma_{ij}
	- a^k \partial_k \gamma_{ij} \nl
	+ \tfrac{1}{2} \omega^{kl} ( M_{kl}^{im} \gamma_{mj} + M_{kl}^{jm} \gamma_{im} ) \,,
\end{split}
\end{equation}
which can be written as
\begin{equation}
\gamma_{ij} \rightarrow \gamma_{ij} - \mathcal{L}_{\delta x^k} \gamma_{ij} \,,
\end{equation}
where $\mathcal{L}$ denotes the Lie-derivative.
[This would of course be valid for any infinitesimal coordinate transformation,
not only for (\ref{eucl}).]
% (We can also write (\ref{eucl}) as a Lie-shift of the coordinate scalars $x^k$, i.e.,
% $\delta x^k = \mathcal{L}_{\delta x^l} x^k$.)

As the Euclidean group is given by its standard representation in the ADMTT gauge,
its generators in phase space are also given by its usual representations,
Eqs.\ (\ref{Ptot}), (\ref{Jtot}), (\ref{Pstd}), and (\ref{Jstd}).
Indeed, the transformation rule for an arbitrary phase space function $A$,
\begin{equation}
	A \rightarrow A + \tfrac{1}{2} \omega^{ij} \{ A , J_{ji} \} + a^i \{ A , P_i \} \,, \label{trans}
\end{equation}
applied to our fundamental variables then reads
\begin{subequations}\label{symtrans}
\begin{align}
\hat{z}_a^i &\rightarrow \hat{z}_a^i + \tfrac{1}{2} \omega^{kl} M_{kl}^{ij} x^j + a^i \,, \\
P_{a i} &\rightarrow P_{a i} + \tfrac{1}{2} \omega^{kl} M_{kl}^{ij} P_{a j} \,, \\
S_{a(i)(j)} &\rightarrow S_{a(i)(j)} + \tfrac{1}{2} \omega^{kl} ( M_{kl}^{im} S_{a(m)(j)} + M_{kl}^{jm} S_{a(i)(m)} ) \,, \\
h^{\text{TT}}_{i j} &\rightarrow h^{\text{TT}}_{i j} - \mathcal{L}_{\delta x^k} h^{\text{TT}}_{i j} \,, \label{htttrans} \\
\pi^{i j \text{TT}}_{\text{can}} &\rightarrow \pi^{i j \text{TT}}_{\text{can}}
	- \mathcal{L}_{\delta x^k} \pi^{i j \text{TT}}_{\text{can}} \,. \label{pitttrans}
\end{align}
\end{subequations}
Thus the generators $J_{ij}$ and $P_i$ in its standard representation give the
transformation induced by (\ref{eucl}) on the fundamental variables,
as expected. In (\ref{htttrans}) and (\ref{pitttrans}) it was used that, e.g.,
$\delta^{\text{TT}kl}_{ij} \mathcal{L}_{\delta x^m} h^{\text{TT}}_{kl} = \mathcal{L}_{\delta x^m} h^{\text{TT}}_{i j}$
which again reflects the compatibility of the ADMTT gauge with the
standard representation of the Euclidean group.

Further, the canonical action given by Eq.\ (51) in \cite{SSepl}
(see also (4.33) in \cite{steinhoff08:spin2}) is invariant under
\begin{equation}
\hat{\lambda}^{[i](j)} \rightarrow \hat{\lambda}^{[i](j)} + \tfrac{1}{2} \omega^{kl} M_{kl}^{jm} \hat{\lambda}^{[i](m)} \,,
\end{equation}
and the transformations (\ref{symtrans}). The corresponding conserved
quantities can be obtained in the standard Noether manner and result as
(\ref{Ptot}) and (\ref{Jtot}) with (\ref{Pstd}) and (\ref{Jstd})
inserted, as expected.

A straightforward (3+1)-split of the Poincar\'e algebra leads to
\begin{align}
\{ P_k , P_i \} &= 0 \,, \\
\{ E , P_i \} &= 0 \,, \\
\{ E , J_{ji} \} &= 0 \,, \\
\{ P_k , J_{ji} \} &= M_{ij}^{kl} P_{l} \,, \\
\{ G^k , J_{ji} \} &= M_{ij}^{kl} G^l \,, \\
\{ J_{kl} , J_{ji} \} &= M_{ij}^{km} J_{ml} + M_{ij}^{lm} J_{km} \,, \\
\{ G^k , P_i \} &= E \delta_{ik} \,, \label{Gtranslat} \\
\{ G^i, G^j \} &= - J_{ij} \,, \label{GG} \\
\{ G^i , E \} &= P_i \,. \label{GE}
\end{align}
In consideration of (\ref{trans}) the first two equations reflect
the translation invariance of $P_j$ and $E$, while the third one
requires $E$ to be a scalar under rotations. Similarly, the next
equations state that $P_k$ and $G^k$ transform as vectors under rotations,
while $J_{kl}$ transforms as a bivector. Equation (\ref{Gtranslat})
means that the center-of-mass $X^k = G^k / E$ has the expected
transformation property under translations, $\{ X^i , P_j \} = \delta_{ij}$.
Thus all except the last two equations are fulfilled by construction if
$J_{ij}$ and $P_i$ are given by its standard representation\footnote{If $G^i$
is determined by an ansatz instead of the integral (\ref{Gint}),
then one should also check (\ref{Gtranslat}).}.
For the calculation in \cite{HS08} the fulfillment of (\ref{GE}) implied
that (\ref{GG}) also holds. However, a generalization of this fact is not known
to the authors.

It is instructive to give a physical interpretation of (\ref{GG})
and (\ref{GE}). Equation (\ref{GE}) can be written as
\begin{equation}
\dot{X}^i = \{ X^i , E \} = \frac{P_i}{E} = \text{const} \,,
\end{equation}
and states that the center-of-mass is moving with constant velocity.
If we define a total spin of the system as
\begin{equation}
S_{ij}^{\text{total}} = J_{ij} - X^i P_j + X^j P_i \,, \label{Stot}
\end{equation}
we get
\begin{align}
\{ X^i, X^j \} &= - \frac{S_{ij}^{\text{total}}}{E^2} \,, \\
\{ S_{ij}^{\text{total}} , P_k \} &= 0 \,, \\
\{ S_{ij}^{\text{total}} , X^k \} &= \frac{P_i S_{kj}^{\text{total}}}{E^2} + \frac{P_j S_{ik}^{\text{total}}}{E^2} \,, \\
\begin{split}
\{ S_{ij}^{\text{total}} , S_{kl}^{\text{total}} \} &=
	\left( \delta_{km} - \frac{P_k P_m}{E^2} \right) M_{ij}^{mn} \hat{S}_{nl}^{\text{total}} \nl
	+ \left( \delta_{lm} - \frac{P_l P_m}{E^2} \right) M_{ij}^{mn} \hat{S}_{kn}^{\text{total}} \,.
\end{split}
\end{align}
These are the Poisson brackets known for the center and spin associated with the SSC $S_{0i}^{\text{total}} = 0$.
Notice that
\begin{equation}
\dot{S}_{ij}^{\text{total}} = \{ S_{ij}^{\text{total}} , E \} = 0 \,.
\end{equation}
One can go over to NW variables by
\begin{align}
\hat{S}_{ij}^{\text{total}} &= S_{ij}^{\text{total}} + \frac{P_i P_k S_{kj}^{\text{total}}}{M (E+M)}
	 + \frac{P_j P_k S_{ik}^{\text{total}}}{M (E+M)} \,, \\
\hat{Z}^i &= X^i - \frac{S_{ij}^{\text{total}} P_j}{M (E + M)} \,,
\end{align}
where $M^2 = E^2 - \vct{P}^2$; see, e.g., \cite{HR74}. This transforms (\ref{Stot}) into
\begin{equation}
J_{ij} = \hat{Z}^i P_j - \hat{Z}^j P_i + \hat{S}_{ij}^{\text{total}} \,, \label{Shtot}
\end{equation}
and finally leads to the standard Poisson brackets
\begin{align}
\{ \hat{Z}^i , P_j \} &= \delta_{ij} \,, \\
\{ \hat{Z}^i, \hat{Z}^j \} &= 0 \,, \label{ZZ} \\
\{ \hat{S}_{ij}^{\text{total}} , \hat{Z}^k \} &= 0 \,, \\
\{ \hat{S}_{ij}^{\text{total}} , \hat{S}_{kl}^{\text{total}} \} &= M_{ij}^{km} \hat{S}_{ml}^{\text{total}} + M_{ij}^{lm} \hat{S}_{km}^{\text{total}} \,.
\end{align}
It still holds $\dot{\hat{Z}}^i = P_i/E = \text{const}$ and $\dot{\hat{S}}_{ij}^{\text{total}} = 0$.
Equation (\ref{GG}) was transformed into (\ref{ZZ}). Thus (\ref{GG}) reflects
the fact that by a straightforward (3+1)-split of the Poincar\'e algebra
one arrives at a center $X^i$ associated with the (noncanonical) SSC $S_{0i}^{\text{total}} = 0$.
% In the center-of-mass frame, i.e., for $P_i = 0$, $X^i$ and
% $Z^i$ coincide, as well as $S_{ij}^{\text{total}}$ and
% $\hat{S}_{ij}^{\text{total}}$.

Notice that the structure of the total angular momentum in
(\ref{Stot}) and (\ref{Shtot}) is the same, but the variables
in (\ref{Stot}) are not standard canonical. It is thus
astonishing that the condition (\ref{Jstd}) uniquely fixes
the canonical spin and position variables in this paper.
This is due to the important additional requirement of having a
constant Euclidean spin-length. In this section, however, even the individual
components of $S_{ij}^{\text{total}}$ and $\hat{S}_{ij}^{\text{total}}$
are constant due to the fact that the total system does not interact, e.g.,
with an external field.

\section{Lie-shift version of the transition to the NW position variable\label{lie}}
For $\delta z^i_a = 0$ one can use a Lie-shift to redefine
the position in $\mathcal{H}^{\text{matter}}$ and $\mathcal{H}^{\text{matter}}_i$, i.e.,
\begin{align}
	\mathcal{H}^{\text{matter}} &\rightarrow \mathcal{H}^{\text{matter}} - \mathcal{L}_{\delta x^{\mu}} \mathcal{H}^{\text{matter}}
		\,,\\
	\mathcal{H}^{\text{matter}}_i &\rightarrow \mathcal{H}^{\text{matter}}_i
		- \mathcal{L}_{\delta x^{\mu}} \mathcal{H}^{\text{matter}}_i \,,
\end{align}
with the shift $\delta x^{\mu}$ on the $a$-th worldline given by
\begin{equation}
	\delta x^{\mu}_a = - \frac{nS^{\mu}_a}{m_a - np_a} \,. \\
\end{equation}
It holds $\delta x_a^0 = 0$ and $p_{ai}\delta x^i_a = 0$. Although spatial derivatives of $p_{a i}$ are not defined
(the linear momentum is only known on the worldline), $p_{ai}$ is treated as a vector field
for the Lie-shift. Therefore we must have
\begin{equation}
	\delta x^k_{a ; i} p_{a k} =
		- \delta x^k_a ( p_{a i ; k} + p_{a k , i} - p_{a i , k} ) = 0 \,, \\
\end{equation}
which precisely cancels the spatial derivatives of $p_{a i}$ introduced by
the Lie-shift. Thus $p_{a i}$, as a vector field, must be parallel transported
to the new worldline without rotation.
The transition to the canonical momentum now has to read
\begin{equation}\label{NWmomShift}
\begin{split}
p_{ai} &= P_{ai} - nS^k_a K_{ik} - \pi^{jk}_{a} \gamma_{jk,i}
	+ \frac{1}{2} \hat{S}_{ajk} \nl
		\times \left[\gamma^{lj}\gamma^{kp}\gamma_{il,p} - \frac{P_{am} P_{aq}}{nP_a(m_a-nP_a)}
			\gamma^{mj}\gamma^{kl}\gamma^{qp}\gamma_{lp,i}\right] \,,
\end{split}
\end{equation}
% \begin{widetext}
% \begin{equation}\label{NWmomShift}
% 	p_{ai} = P_{ai} - nS^k_a K_{ik} - \pi^{jk}_{a} \gamma_{jk,i}
% 		+ \frac{1}{2} \left[\gamma^{lj}\gamma^{kp}\gamma_{il,p} - \frac{P_{am} P_{aq}}{nP_a(m_a-nP_a)}
% 			\gamma^{mj}\gamma^{kl}\gamma^{qp}\gamma_{lp,i}\right]\hat{S}_{ajk} \,,
% \end{equation}
% \end{widetext}
in order to satisfy Eq.\ (\ref{defP}). For $\pi^{ij}_{a} = 0$ the momentum redefinition
from \cite{steinhoff08:spin2} is obtained.
Further, this leads to the results (\ref{NWcham}) and (\ref{NWcmom}) for the case $\delta z^i_a = 0$.
However, this formulation is not so useful for variable redefinitions in an action approach.
Notice that (\ref{NWmomShift}) is missing a term when compared to (\ref{NWmom}), however, here $p_{ai}$
is treated as a vector field and is not held constant for the redefinition
of the position.

\section{Center-of-mass frame\label{COM}}
The center-of-mass frame is defined here by the condition that
the total linear momentum and the center-of-mass vector vanish,
i.e., $P_i=G^i=0$. As $P_i$ is conserved and $G^i = P_i t + K^i$ with
$K^i = \text{const}$ (cf. Sec.\ \ref{pgen}) this is indeed a consistent set of constraints
that can be imposed on the phase space for all times $t$. We will restrict to two
objects here. Then $P_i=0$ results in $\vct{P}_1 = - \vct{P}_2$
at the considered order.
The leading order terms of $G^i$ read
\begin{equation}
\vct{G} = m_1 \hat{\vct{z}}_1 + m_2 \hat{\vct{z}}_2 + \frac{\vct{P}_1 \times \vct{S}_1}{2 m_1}
	+ \frac{\vct{P}_2 \times \vct{S}_2}{2 m_2} \,.
\end{equation}
From $\vct{G}=0$ follows
\begin{equation}
	\hat{\vct{z}}_1 = \frac{\mu}{m_1} \vct{r}_{12} - \frac{\vct{p} \times \vct{S}_1}{2 m_1 M}
		+ \frac{\vct{p} \times \vct{S}_2}{2 m_2 M} \,,
\end{equation}
with $\mu = \nu M$, and similar for $\hat{z}^i_2$.
The leading order relation between canonical momentum and velocity reads
\begin{equation}
	\dot{\hat{z}}^i_1 = \frac{P_{1 i}}{m_1} - \left( \frac{3 m_2}{2 m_1} S_{1(i)(j)} + 2 S_{2(i)(j)} \right) \frac{n_{12}^j}{r^2} \,,
\end{equation}
and similar for $\dot{\hat{z}}^i_2$. It follows
\begin{equation}
	\vct{v} = \frac{\vct{p}}{\mu} - \frac{\vct{n}_{12} \times \vct{S}_1}{r^2} \left( 2 + \frac{m_2}{m_1} \right)
		- \frac{\vct{n}_{12} \times \vct{S}_2}{r^2} \left( 2 + \frac{m_1}{m_2} \right) \,.
\end{equation}
The reduced phase space, where $\hat{\vct{z}}_1$, $\hat{\vct{z}}_2$, $\vct{P}_1$, and $\vct{P}_2$
are replaced by $\vct{r}_{12}$ and $\vct{p}$, is then subject to the Poisson bracket
\begin{equation}
\{ r_{12}^i , p_j \} = \delta_{ij} \,.
\end{equation}
This can also be seen as a Dirac-bracket following from $P_i=G^i=0$.

% PRD version
%\bibliographystyle{h-physrev5}
% arXiv version
%\bibliographystyle{utphys}
% local version
%\bibliography{ref}
% upload version, arXiv and PRD
\providecommand{\href}[2]{#2}\begingroup\raggedright\endgroup

\end{document}